\documentclass[journal=jctcce,manuscript=article]{achemso}

\usepackage{chemformula} 
\usepackage[T1]{fontenc} 
\usepackage{titlesec}
\usepackage{amsmath}
\usepackage{amssymb}
\usepackage{braket}
\usepackage{mathtools}
\usepackage{multirow}
\usepackage{enumitem}
\usepackage{float}
\usepackage{graphicx}
\usepackage[caption=false]{subfig}
\usepackage{algorithm}
\usepackage{algpseudocode} 

\author{Qiujiang Liang}
\author{Jun Yang}
\email{juny@hku.hk}
\affiliation[The University of Hong Kong]
{Department of Chemistry, The University of Hong Kong, 
Hong Kong SAR, P.R. China}

\title[]
  {
Third-order many-body expansion of OSV-MP2 wavefunction for low-order scaling
analytical gradient computation
   }

\graphicspath{{./figures/}}
\begin{document}

\begin{abstract}

We present a many-body expansion (MBE) formulation and implementation for
efficient computation of analytical energy gradients from the
orbital-specific-virtual second-order M{\o}llet-Plesset perturbation theory
(OSV-MP2) based on our earlier work (Zhou et al. \textit{J.  Chem.  Theory
Comput.} \textbf{2020}, \textit{16}, 196--210). The third-order MBE(3) expansion
of OSV-MP2 amplitudes and density matrices was developed to adopt the
orbital-specific clustering and long-range  termination schemes, which avoids
term-by-term differentiations of the MBE energy bodies. We achieve better
efficiency by exploiting the algorithmic sparsity that allows to prune out
insignificant fitting integrals and OSV relaxations.  With these approximations,
the present implementation is benchmarked on a range of molecules that show an
economic scaling in the linear and quadratic regimes for computing
MBE(3)-OSV-MP2 amplitude and gradient equations, respectively, and yields normal
accuracy comparable to the original OSV-MP2 results. The MPI-3-based parallelism
through shared memory one-sided communication is further developed for improving
parallel scalability and memory accessibility by sorting the MBE(3) orbital
clusters into independent tasks that are distributed on multiple processes across
many nodes, supporting both global and local data locations in which selected
MBE(3)-OSV-MP2 intermediates of different sizes are distinguished and
accordingly placed. The accuracy and efficiency level of our MBE(3)-OSV-MP2
analytical gradient implementation is finally illustrated in two applications:
we show that the subtle coordination structure differences of mechanically
interlocked Cu-catenane complexes can be distinguished when tuning ligand
lengths; and the porphycene molecular dynamics reveals the emergence of the
vibrational signature arising from softened N-H stretching associated with
hydrogen transfer, using an MP2 level of electron correlation and classical
nuclei for the first time.

\end{abstract}
\section{INTRODUCTION}

Correlated post-Hartree-Fock (post-HF) methods are being advanced rapidly in the
past decade for enabling energy computation of large molecules with
systematically controlled accuracy. The strategies for ameliorating post-HF
complexities are typically based on two main streams: local correlation methods in
which the locality\cite{pulay1983localizability} or
near-sightedness\cite{kohn1996density} of electrons is explored within full
system in different ways, and fragmentation- or subdomain-based idea of many
variants in which the original formidable problem is broken into many smaller
pieces of amenable subproblems. The first stream approaches the full solution by
compressing the cluster operators of the entire system in various low-order
scaling post-HF methods, predominantly the popular second-order
M{\o}llet-Plesset (MP2) perturbation
\cite{maslen1998non,ayala1999linear,lee2000closely,doser2009linear,yang2011tensor,kurashige2012optimization,werner2015scalable}
and coupled-cluster (CC) theory
\cite{hampel1996local,schutz2000local,schutz2001low,schutz2002low,schutz2002new,subotnik2005local,auer2006dynamically,subotnik2008limits,neese2009efficient,werner2011efficient,yang2012orbital,schutz2013orbital,riplinger2013efficient,riplinger2013natural},
in the  local frameworks such as projected atomic orbitals (PAO),
\cite{hampel1996local,schutz2000local,schutz2001low,schutz2002low,schutz2002new,werner2011efficient}
pair-nature-orbitals
(PNOs)\cite{meyer1971ionization,ahlrichs1975pno,neese2009efficient,pinski2018communication,pinski2019analytical,stoychev2021dlpno}
and orbital-specific-virtuals (OSVs)
\cite{yang2011tensor,kurashige2012optimization,yang2012orbital,schutz2013orbital,zhou2019complete}.
The second stream seeks and combines many subsystem solutions which aim to
approximate the original full solution of the energy via truncated $n$-order
many-body expansion (MBE($n$))
\begin{equation}
E = \sum_I^N E_I+\sum_{I>J}^N\Delta E_{IJ}+\sum_{I>J>K}^N \Delta E_{IJK}+\cdots,
  \label{eq:mbe1}
\end{equation}
with a myriad of prescriptions for the two-, three-, \dots, $n$-body subsystems
and energy corrections in different versions of MBE, sometimes mutually
inclusive, when integrated with correlated wavefunction methods, including the
divide-and-conquer \cite{forner1985coupled,li2004divide,kobayashi2006second},
the incremental scheme,
\cite{stoll1992correlation,stoll1992on,doll1995correlation,kalvoda1998ab,friedrich2007fully,friedrich2008implementation,friedrich2009fully,kallay2015linear,nagy2016integral}
the natural linear scaling method\cite{flocke2004natural,hughes2008natural}, the
cluster-in-molecule,
\cite{li2002linear,li2006efficient,li2009local,rolik2011general,rolik2013efficient}
the fragment molecular orbital,
\cite{kitaura1999fragment,nagata2011mathematical,fedorov2012exploring,gordon2012fragmentation}
the embedded
MBE\cite{hirata2005fast,dahlke2007electrostatically,hirata2008fast,fujita2011fragment,bygrave2012embedded}
and several others
\cite{wen2011accurate,zalesny2011linear,gordon2017fragmentation,liu2019energy,herbert2019fantasy}.

Both streams have been intensively developed in recent years for correlation
treatments and become closely related, for targeting previously difficult
systems for which energies now can be computed within reasonable accuracy and
time, as demonstrated for thousand-atom
MP2\cite{mochizuki2008large,doser2009linear,nagy2016integral,kjaergaard2017massively}
and hundred-atom CC
\cite{riplinger2013efficient,riplinger2016sparse,guo2018comparison} benchmark
works.
Moreover, many structural and spectroscopic properties can be also cast as
energy derivatives of the MBE eq~\ref{eq:mbe1} with respect to a perturbation, 
\begin{equation}
  \frac{dE}{d\lambda} = \sum_I^N \frac{dE_I}{d\lambda}+\sum_{I>J}^N\frac{d\Delta
  E_{IJ}}{d\lambda}+\sum_{I>J>K}^N \frac{d\Delta E_{IJK}}{d\lambda}+\cdots.
  \label{eq:mbe2}
\end{equation}
This underlies very promising protocol for computing the properties of extended
systems at correlated wavefunction level from MP2 to CC using term-by-term
numerical or analytical differentiations of eq~\ref{eq:mbe2} up to tractable
orders. In recent years, successful applications have been highlighted in a
number of \textit{ab-initio} problems involving electric-field derivatives for
electric moments\cite{friedrich2009implementation,fiedler2016molecular},
polarizabilities\cite{yang2007evaluation,friedrich2015incremental} and
vibrational spectra\cite{he2012second,hirata2014ab,sahu2015accurate}, and
nuclear gradients for geometry optimizations
\cite{hirata2008fast,nagata2011analytic,kristensen2012molecular,bykov2016molecular,ni2019analytical}
and molecular dynamics (MD) simulations
\cite{mochizuki2011fragment,willow2015ab,spura2015fly,li2016hybrid,
pruitt2016importance,haycraft2017efficient,liu2018hydrogen,pham2020development}.
A tremendous variety of these MBE methods for both energies and properties
features the fragmentation schemes in which the subsystems are formed by
properly and explicitly cutting macromolecule into overlapping or
non-overlapping atomic
fragments\cite{mayhall2011molecules,richard2012generalized} prior to post-HF.
Alternatively interests also focus on designating the subsystems as tractable
``\textit{bodies}'' by grouping orbital domains based on the starting HF
wavefunction of the supersystem.  

Although MBE provides a general skeleton to integrate with arbitrary electronic
structure methods, it becomes practically tractable only if the sequences of
eq~\ref{eq:mbe1} for energies and eq~\ref{eq:mbe2} for observables are well
converged and terminated at low expansion orders, for instance, $n\le3$ aiming
for high accuracy. When we consider the cumulative CPU time $t(n)$ as measurement
of the needed hardware resource for computing the MBE($n$) up to the order $n$,
apparently $t(n) \approx m_1 \bar t_1 + m_2 \bar t_2  + \cdots + m_n \bar t_n$
depends on a few factors for a given macromolecule: (1) the number of
independent $i$-body subsystems ($m_i$); (2) the size of individual subsystem
($N_i$) that needs the average CPU time ($\bar t_i\sim {\cal{O}}(N_i^p)$, e.g.,
roughly $p=5$ for canonical MP2 and $p=6$ for CCSD); (3) and the orbital
topology belonging to each subsystem which is usually determined at the
mean-field level and affects the post-HF MBE($n$) convergence at an expansion
order $n$.  While it is obvious that such a large number of independent
computations must be leveraged in efficient massive parallelism, vast hardware
costs can be saved in computations for which both $m_i$ and $t_i$ increase only
moderately (e.g., ${\cal{O}}(N)$) with size of macromolecules. By compressing
$m_i$, the key idea is to compute only a subset of important MBE terms
explicitly, usually in the presence of classical electrostatic and/or
approximate dispersion potentials that implicitly fold the corrections from the
long-range and high-order nonadditive many-body terms. On the other hand, as the
sizes $N_i$ of subsystems control the cost of each correlated MBE computation,
one could think of computations of lower expense for smaller subsystems, which
however often pose difficulty in converging MBE errors and may involve an
excessively large number of subsystems.  It is therefore desirable to combine
the MBE with low-scaling local correlation methods by creating and computing the
subsystems of compact orbital topology.  In recent years, this strategy has been
carefully examined in connection with PAO/PNO/OSV virtual space representation
for computing energies of large molecular
clusters\cite{fiedler2017combining,guo2018comparison} and also applied to
molecular crystals.
\cite{yang2014ab,beran2016modeling,wang2019cluster,hansen2020representation}

In this work, we present an MBE extension of these essential ideas to include
energy derivatives that will be rewarding macromolecules based on the local
OSV-MP2 analytical gradient formulation we recently
developed\cite{zhou2019complete}. It was shown that the formal scaling of
canonical MP2 gradient computation was lowered by about 2--3 orders of
magnitude using OSV-MP2.  Nevertheless, OSV-MP2 gradient computation is still a
resource-intensive task compared to energy.  One can envisage that when each MBE
1-body (1b) subsystem is inevitably very large, the large 1b subsystems create
superlarge 2-body (2b) and 3-body (3b) subsystems which are prohibitively
expensive even for low-scaling methods.  By realizing that the OSV ans\"atz
provides the inherently compact representation for virtual space that is adapted
to a single molecular orbital (MO), this orbital-specific nature makes OSV a
convenient choice for deploying an MBE(3) sequence in which very small pieces of
OSV-MP2 analytical gradient computations can be performed on minimum subsystems
in the spirit of energy incremental
scheme\cite{stoll1992correlation,stoll1992on}, i.e., by correlating only
individual local MO (LMO) for 1b, LMO pair $ij$ for 2b, and LMO triple
$ijk$ for 3b at a time, and keeping other electrons inactive.  As such, the
OSV virtual domain becomes optimal to correlate 1b LMO (i.e., a pair of
electrons), and the union of 1b OSVs creates the local domains specific to 2b
and 3b LMOs, respectively.  We will show that this MBE($3$)-OSV-MP2 approach
already converges OSV-MP2 gradients very well for structure and molecular
dynamics (MD) simulations compared to canonical results, without resorting to
auxiliary embeddings. This avoids the well known complications of implementing
and computing the analytical gradient arising from each subsystem's response of
nonfully variational embedding potential due to the changes of other subsystems.
More importantly for complex molecules, $m_1$ for 1b subsystems exhibits a
natural linear growth with respect to macromolecular size, and $m_2$ and $m_3$
increase as ${\cal{O}}(N)$ when the intrinsic sparsity within 2b and 3b domains
is exploited based on the OSV-based metric.

The remaining discussions are organized as follows. Section~\ref{sec:review}
briefly reviews and reformulates the OSV-MP2 analytical gradient theory, and
section~\ref{sec:mbegrad} discusses the third-order MBE(3)-OSV-MP2 algorithm and
implementation details for parallel gradient computations based on MPI-3
standard. All integrals and their geometric derivatives are computed using the
quantum chemistry program package \texttt{PYSCF}\cite{sun2020recent}.
Sections~\ref{sec:mbeerror} and~\ref{sec:mbeorder} discuss the performance of
the MBE(3)-OSV-MP2 implementation by assessing the accuracy, the origin of
errors and the parallel efficiency for computing molecular structures and
dynamical properties.  Finally, in sections~\ref{sec:catenane}
and~\ref{sec:porphycene}, we illustrate two short MBE(3)-OSV-MP2 applications in
determining the subtle structural changes of Cu-Catenane interlocking complex
with varying ligand lengths, as well as molecular dynamics simulation showing
protonic tautomerization in porphycene molecule.

\section{THEORY AND IMPLEMENTATION}

\subsection{\label{sec:review}Review of OSV-MP2 Gradient Theory}

We briefly discuss a reformulation of OSV-MP2 analytical gradient
theory\cite{zhou2019complete}, which is essential to the implementation of its
MBE extension. We adopt the following convention for noting orbitals:
$i,j,k,\cdots$ and $a,b,c,\cdots$ represent the occupied LMOs ($\mathbf{C}^o$)
and canonical virtual MOs $\mathbf{C}^v$, respectively;
$\bar\mu_k,\bar\nu_k,\bar\xi_k,\cdots$ refer to a set of OSV orbitals specific
to the occupied LMO $k$; $p,q,r,\cdots$ and $\alpha,\beta,\cdots$ pertain to
generic indices of MOs and atomic orbitals (AOs), respectively. For brevity, the
matrix trace operation is denoted by the bra-ket $\braket{\cdots}$. All matrices
and elements are signified by bold and italic letters, respectively.

The OSV-MP2 correlation energy $E_c$ is computed according to the
orbital-invariant Hylleraas Lagrangian,
\begin{equation}
 E_c = \sum_{ij}\braket{\mathbf{K}_{(ij,ij)}\overline{\mathbf{T}}_{(ij,ij)}}
    + \braket{\mathbf{R}_{(ij,ij)}\overline{\mathbf{T}}_{(ij,ij)}} \label{eq:ec}
\end{equation}
with
$\overline{\mathbf{T}}_{(ij,ij)}=2\mathbf{T}_{(ij,ij)}-\mathbf{T}_{(ij,ij)}^\dagger$.
The Hylleraas energy minimization with respect to the pair amplitudes
$\mathbf{T}_{(ij,ij)}$ yields a set of residual equations that must be solved
iteratively in the OSV basis,
\begin{equation}
  \mathbf{R}_{(ij,ij)} = \mathbf{K}_{(ij,ij)} + \sum_k \mathbf{S}_{(ij,ik)} 
  \mathbf{T}_{(ik,ik)} [\delta_{kj} \mathbf{F}_{(ik,ij)}-f_{kj}\mathbf{S}_{(ik,ij)}]
  +
[\delta_{ik}\mathbf{F}_{(ij,kj)}-f_{ik}\mathbf{S}_{(ij,kj)}]\mathbf{T}_{(kj,kj)}\mathbf{S}_{(kj,ij)}.
   \label{eq:rij}
\end{equation}
The pair amplitudes $\mathbf{T}_{(ij,ij)}$, the residual equations
$\mathbf{R}_{(ij,ij)}$ and the relevant quantities $\mathbf{A}_{(ij,kl)}$ are
manipulated and stored in the OSV basis of highly compressed dimension, by
associating each set of compact OSV orbitals $\{\bar{\mu}_{k}\}$ with individual
occupied orbital $k$ through one-index transformation from canonical virtuals
$\{a\}$
\begin{equation}
  \ket{\bar\mu_k} = \sum_a Q_{a\bar\mu}^{k} \ket a,\label{eq:osv}
\end{equation}
where the $k$-specific transformation matrix $\mathbf{Q}_{k}$ is determined by
taking the orthonormal eigenvector of the MP2 diagonal pair amplitudes
$\mathbf{T}_{kk}$ for each $k$,
\begin{equation}
 \left[ \mathbf{Q}_k^\dagger \mathbf{T}_{kk} \mathbf{Q}_{k} \right]_{\bar\mu_k\bar\nu_k}
  = \omega_{\bar\mu_k} \delta_{\bar\mu\bar\nu}, \label{eq:svd}
\end{equation}
with the orthonormality $\mathbf{Q}_k^\dagger \mathbf{Q}_{k} = \mathbf{1}$.
The elements 
$[\mathbf{T}_{kk}]_{ab} = \frac{(k a\rvert k b)}{f_{aa}+f_{bb}-2f_{kk}}$
are computed using the diagonal elements $f_{kk}, f_{aa}, f_{bb}$ of the Fock
matrix. The level of compactness of OSV space is controlled by the column
dimension of the vectors $\mathbf{Q}_{k}$ having eigenvalues
$\omega_{\bar{\mu}_{k}}$ greater than a cut-off $l_{\text{osv}}$.

In eq~\ref{eq:rij}, we adopt $\mathbf{A}_{(ij,kl)}$ to represent a generic composite matrix in
the OSV-concatenated pair domain that must be assembled between
$\{\bar\mu_i,\bar\nu_j\}$ and $\{\bar\sigma_k,\bar\xi_l\}$ elements.
For instance, $\mathbf{K}_{(ij,ij)}$
denotes the OSV two-electron integral assembled from the composition of
$(i \bar\mu_i \rvert j \bar\nu_i)$, $(i \bar\mu_i \rvert j \bar\xi_j)$, 
$(i \bar\sigma_j \rvert j \bar\nu_i)$ and $(i \bar\sigma_j \rvert j \bar\xi_j)$.
$\mathbf{A}_{(ij,kl)}$ can be however conveniently expressed as a projection of
$\mathbf{A}$ from the canonical virtual MOs to OSVs basis and is self-adjoint
$\mathbf{A}_{(ij,kl)}^{\dagger}=\mathbf{A}_{(kl,ij)}\label{eq:aadjoint}$,
\begin{equation}
 \mathbf{A}_{(ij,kl)} = 
 \left( 
  {\begin{array}{c} 
    \mathbf{Q}_i^\dagger\\\mathbf{Q}_j^\dagger 
   \end{array}}
 \right)
  \mathbf{A}
 \left( 
  {\begin{array}{cc} 
    \mathbf{Q}_k & \mathbf{Q}_l
   \end{array}}
 \right). 
\label{eq:aijkl}
\end{equation}
In the OSV-based analytical gradient theory, the OSV derivative of $
\mathbf{A}_{(ij,kl)}$ is needed and formulated by employing the OSV relaxation
matrix $\mathbf{O}_k^\lambda$ with respect to a perturbation $\lambda$,
\begin{equation}
 \mathbf{A}_{(ij,kl)}^{\{\lambda\}} = 
 \left( 
  {\begin{array}{cc} 
    \mathbf{O}_i^{\dagger\lambda} & \mathbf{0}\\\mathbf{0} & \mathbf{O}_j^{\dagger\lambda}
   \end{array}}
 \right)
  \mathbf{A}_{(ij,kl)}^0 + \mathbf{A}_{(ij,kl)}^0
 \left( 
  {\begin{array}{cc} 
    \mathbf{O}_k^{\lambda} & \mathbf{0}\\ \mathbf{0} & \mathbf{O}_l^{\lambda}
   \end{array}}
 \right),
 \label{eq:aijklgrad}
\end{equation}
with the curly brackets $\{\}$ specifying the derivatives of OSVs. Only the
off-diagonal block of $\mathbf{O}_k^{\lambda}$ between the kept and discarded
OSV spaces is needed for accounting effective OSV relaxations, which is cast as
perturbed nondegenerate eigenvalue problem\cite{zhou2019complete}, requiring the
first derivative of the diagonal pair amplitudes $\mathbf{T}_{kk}$. 

The OSV-MP2 correlation energy
$E_{c}[\alpha,\mathbf{C}^o,\mathbf{Q}_k,\mathbf{T}_{(ij,ij)}]$ can be
viewed as a function of a string of variables: atomic orbitals
$\alpha,\beta,\cdots$, LMOs $\mathbf{C}^o$, OSVs $\mathbf{Q}_k$ and pair
amplitudes $\mathbf{T}_{(ij,ij)}$.  As the amplitudes $\mathbf{T}_{(ij,ij)}$ are
variational to Hylleraas $E_{c}$ and make no contribution, the OSV-MP2 energy derivative
$E^{\lambda}_{c}=\frac{dE_{c}}{d\lambda}$ can be computed according to the
relaxation contributions merely from OSVs ($E^{\{\lambda\}}_c$), LMOs
($E^{[\lambda]}_c$) and AOs ($E^{(\lambda)}_c$) with respect to a geometric
perturbation $\lambda$,
\begin{equation}
    E^{\lambda}_{c} = E^{\{\lambda\}}_{c} + E^{[\lambda]}_{c} + E^{(\lambda)}_{c}.
\end{equation}
The OSV-specific energy gradient $E^{\{\lambda\}}_c$ results from the OSV
responses of both the residual equations collected in pair intermediates
$\mathbf{M}_{ij}$, and the OSV-based integrals  in form of
$\mathbf{A}_{(ij,ij)}$ for the integrals $\mathbf{K}_{(ij,ij)}$,
$\mathbf{F}_{(ij,ij)}$ and $\mathbf{S}_{(ij,ij)}$.  The MO-specific
$E^{[\lambda]}_{c}$ arises from the LMO relaxation jointly determined by the
geometric responses of canonical MOs and the localization function, requiring
the solution to the coupled-perturbed HF and the coupled-perturbed localization
equation\cite{el1998analytical}, respectively. In our implementation, their
contributions are merged into the OSV-based Z-vector equation. The AO-specific
$E^{(\lambda)}_c$ simply evaluates the Hylleraas energy expression of
eq~\ref{eq:ec} in terms of two- and one-electron AO derivative integrals, the
occupied-occupied block ($D_{ij}$) and OSV-OSV block ($\mathbf{D}_{(ij,ij)}$) of
the unrelaxed density matrices.

Combining all three gradient contributions and using the resolution-of-identity
(RI) integral approximation, the total OSV-MP2 energy gradient is reformulated
in terms of a variety of density matrices together with the AO-derivatives of
Fock (${F}_{\alpha\beta}^{(\lambda)}$), overlap
(${S}_{\alpha\beta}^{(\lambda)}$), half-transformed 3-center-2-electron (3c2e)
integral ($\mathbf{J}_i^{(\lambda)}$) matrices, 
\begin{equation}
 E_c^{\lambda} = \braket{[\mathbf{\check D} 
       -\mathbf{C}^o\mathbf{Z}^\dagger \mathbf{C}^{v\dagger} ]\mathbf{F^{(\lambda)}}}
       - \braket{[\mathbf{\check D}' + \mathbf{C}^o\mathbf{Z}^\dagger\mathbf{C}^{v\dagger}
       + \frac{1}{2}\mathbf{C}^o\braket{\mathbf{Z^\dagger A}}\mathbf{C}^{o\dagger}]
        \mathbf{S^{(\lambda)}} }
      + 4\braket{\sum_i \mathbf{P}^v\mathbf{Y}_i^\dagger\mathbf{J}_i^{(\lambda)}}.
\label{eq:ec_grad}
\end{equation}
As seen here, the first two trace terms account for the effective one-electron
response and the last term for the two-electron response. The one-electron
response results eventually from the internal-external orbital rotation by
OSV-MP2 Z-vector $\mathbf{Z}$ as well as collective density matrices
$\mathbf{\check D}$ and $\mathbf{\check D}'$ in AO basis.  $\mathbf{\check D}$
collects the direct sum of the OSV overlap-weighted external and internal
density matrices ($\mathbf{D}^v$ and $\mathbf{D}^o$) in the MO basis,
\begin{eqnarray}
   \mathbf{\check{D}} &=& 2 \left ( \mathbf{C}^o~\mathbf{C}^v \right) 
   \left[ \mathbf{D}^o \oplus \mathbf{D}^v \right]
   \left ( \mathbf{C}^o~\mathbf{C}^v \right)^\dagger \label{eq:dm_s}
\end{eqnarray}
with overlap-weighted density matrices in the MO basis,
\begin{eqnarray}
    \mathbf{D}^v &=& \sum_{ij}
     \left(\mathbf{Q}_i~\mathbf{Q}_j\right)
            \mathbf{D}_{(ij,ij)} 
     \left(\mathbf{Q}_i~\mathbf{Q}_j\right)^\dagger +\sum_{ij} \mathbf{T}_{ii}\mathbf{X}_{ij},
\label{eq:dv} \\
  {D}^o_{ij} &=& - D_{ij} - \delta_{ij}\braket{\mathbf{T}_{i i} \sum_{k}\mathbf{X}_{ik}}.
\label{eq:do} 
\end{eqnarray}
Above, $\mathbf{D}^v$ and $\mathbf{D}^o$ are composed of the unrelaxed and
relaxed contributions. $\mathbf{X}_{ij}$ is an important intermediate
resembling the relaxed amplitudes in the MO basis, which accounts for the
geometric OSV relaxations  of two-electron integrals and residuals, computed in
terms of $\mathbf{N}_{ij}$ for the pair $ij$, 
\begin{equation}
\mathbf{N}_{ij} = \overline{\mathbf{T}}_{(ij,ij)}\mathbf{K}_{(ij,ij)} 
         + \mathbf{D}_{(ij,ij)}\mathbf{F}_{(ij,ij)} + \mathbf{D}_{(ij,ij)}^\prime \mathbf{S}_{(ij,ij)}
      - \sum_k \left[f_{jk}\mathbf{D}_{(ij,ik)}\mathbf{S}_{(ik,ij)}+f_{ik}\mathbf{D}_{(ij,kj)}\mathbf{S}_{(kj,ij)}\right],
\label{eq:nij}
\end{equation}
where the overlap- ($\mathbf{D}_{(i j, k l)}$) and energy-weighted
($\mathbf{D}_{(i j, k l)}^\prime$) density matrices in the OSV basis are,
\begin{eqnarray}
\mathbf{D}_{(ij,kl)}&=&\frac{1}{2}\left[\overline{\mathbf{T}}_{(ij,ij)}
\mathbf{S}_{(ij,kl)} \mathbf{T}_{(kl,kl)}+\overline{\mathbf{T}}_{(ij,ij)}^{\dagger}
\mathbf{S}_{(ij,kl)} \mathbf{T}_{(kl,kl)}^{\dagger}\right],
\\
\mathbf{D}_{(ij,kl)}^{\prime}&=&\frac{1}{2}\left[\overline{\mathbf{T}}_{(ij,ij)}
\mathbf{F}_{(ij,kl)} \mathbf{T}_{(kl,kl)}+\overline{\mathbf{T}}_{(ij,ij)}^{\dagger} 
\mathbf{F}_{(ij,kl)} \mathbf{T}_{(kl,k l)}^{\dagger}\right].
\end{eqnarray}
Analogously, $\mathbf{\check{D}'}$ collects the OSV energy-weighted external
contribution ($\mathbf{D}'^v$), the internal contribution ($\mathbf{D}'^o$), the
HF occupied ($\mathbf{P}^o$) and virtual
($\mathbf{P}^v$) density matrices for reducing two-electron terms,
\begin{eqnarray}
 \mathbf{\check D}' &=& 2( \mathbf{C}^o~\mathbf{C}^v ) 
   [(\mathbf{D}'^o-\braket{\mathbf{J}_i^\dagger\mathbf{Y}_j\mathbf{P}^v}) 
   \oplus (\mathbf{D}'^v -\mathbf{\Lambda}\mathbf{\mathcal{A}})]
\left ( \mathbf{C}^o~\mathbf{C}^v \right)^\dagger
\nonumber \\
 &&     + 2\sum_i \mathbf{P}^v(
      \mathbf{J}_i^\dagger\mathbf{Y}_i\mathbf{P}^v
    - \mathbf{Y}_i^\dagger\mathbf{J}_i\mathbf{P}^o
      )
\label{eq:dm_e}
\end{eqnarray}
with energy-weighted density matrices in the MO basis,
\begin{eqnarray}
 \mathbf{D}'^v &=& \frac{\mathbf{F}^v\mathbf{D}^v+\mathbf{D}^v\mathbf{F}^v}{2}, \\
 {D}'^o_{ij}&=& - \sum_k f_{ik}D_{ki}-f_{ij}\braket{\mathbf{T}_{i i}\sum_{j}\mathbf{X}_{ij}}.
\end{eqnarray}
The two-electron response associated with $\mathbf{J}_i^{(\lambda)}$ in
eq~\ref{eq:ec_grad} is driven by the intermediate $\mathbf{Y}_i$ 
\begin{equation}
  \mathbf{Y}_i =  \sum_{j}  \mathbf{J}_{j}\left(\mathbf{\check Q}_i~\mathbf{\check Q}_j \right)\
  \overline{\mathbf{T}}_{(ij,ij)}
   \left(\mathbf{\check Q}_i~\mathbf{\check Q}_j \right)^\dagger+\mathbf{J}_i\mathbf{X}_{ij}.
  \label{eq:ec2eY}
\end{equation}

The remaining vector $\mathbf{\Lambda}$ of eq~\ref{eq:dm_e} for Pipek-Mezey
localization constraint and the $\mathbf{Z}$ of eq~\ref{eq:ec_grad} for
internal-external rotation is the respective solution to the linear
coupled-perturbed localization and OSV Z-vector equation,
\begin{eqnarray}
 \mathbf{\mathcal{C}}^\dagger \mathbf{\Lambda}& = & \mathbf{\Gamma}^\dagger,\label {eq:cpl}\\
 \mathbf{A}^\dagger \mathbf{Z} &=& \mathbf{W}.
 \label{eq:zvec}
\end{eqnarray}
$\mathbf{\Gamma}$ and $\mathbf{W}$ on the right are composed of the elements
below, respectively,
\begin{eqnarray}
{\Gamma}_{ij} &=& \mathbf{D}'^o_{ij} - \braket{\mathbf{J}^{\dagger}_i\mathbf{Y}_j\mathbf{P}^v},\\
{W}_{ai} & = & \braket{\mathbf{P}^v\mathbf{Y}_i^{\dagger}\mathbf{J}_a}
           + \sum_j[\mathbf{Y}_j^\dagger\mathbf{J}_j\mathbf{C}^o]_{ai}
           + 2\sum_{kl}{\Lambda}_{kl}\mathcal{B}_{kl,ai}.\label{eq:sourcew}
\end{eqnarray}
The two-electron integrals $\mathbf{A}$ are evaluated with RI approximation,
\begin{equation}
  A_{ai,bj}=\delta_{ap}\delta_{ij}(f_{aa}-f_{ii})+4(ai\rvert bj)-(ap\rvert ij)-(aj\rvert bi).
\label{eq:zveca}
\end{equation}
More details of these intermediate quantities in
eqs~\ref{eq:ec_grad}--\ref{eq:zvec} can be found in our previous
work\cite{zhou2019complete}.

\subsection{\label{sec:mbegrad}MBE(3)-OSV-MP2 Gradient Method and Implementation}

\subsubsection{MBE(3) partitioning, clustering and expansion}

The ability to leverage the OSV-MP2 analytical gradient algorithm for efficient
large scale computations is based on an MBE partitioning in which the $N$ LMOs
from the HF solution of macromolecule is divided into $m_1=N$ 1b subsystems.
Each $i$-th 1b subsystem is coined a \textit{1b cluster}, which constitutes a
small number of prescribed OSVs $\{\bar\mu_i\}$ that become specific to this 1b
cluster by the nature of the generation of OSVs, and correlates a pair of
electrons within the excitation manifolds $i \to \{\bar\mu_i\}$. As such,
the size of each 1b cluster remains minimum, enabling very small OSV-MP2
gradient computation, and the length of all $N$ 1b clusters grows naturally
linearly with sizes of macromolecule. 

The union of two 1b clusters makes a \textit{2b cluster} in which two electron
pairs are correlated in a combined set of the 1b excitation manifolds
$(i,j)\to \{\bar\mu_i\}\cup\{\bar\nu_j\}$ specific to this pair $ij$.
Although the generic length of 2b clusters scales as $N(N+1)/2$, due to the
locality of electron correlations which decrease rapidly with distance, the
contributions to OSV-MP2 gradients from many weak 2b clusters which are made of
relatively remote 1b clusters can be accurately approximated with negligible
costs, as compared to that for strong 2b clusters.  As a result, a linear growth
of the number of the strong 2b clusters can be anticipated as well, which we
will discuss in further section. 

In contrast to canonical MP2 theory which deals with canonical 2b interactions
rigorously, the OSV-MP2 method demands at least indirect 3b corrections to the
local 2b interactions in the presence of other LMOs $k$, as clearly indicated in
the residual eq~\ref{eq:rij}. The \textit{3b clusters} are composed of an
incremental union of three 1b clusters for the excitation manifolds
$(i,j,k)\to \{\bar\mu_i\}\cup\{\bar\nu_j\}\cup\{\bar\sigma_k\}$.
Nevertheless, these 3b corrections must encompass extremely strong pairwise
interactions that are simultaneously present amongst $ijk$ LMOs, and higher
MBE orders than 3b can be also safely neglected for OSV-MP2 gradients, which
defines the MBE(3)-OSV-MP2 ans\"atz that terminates the MBE series at the
third-order.  As a result, a substantial amount of 3b clusters can be discarded
for 3b contributions, leading to a linear growth of the 3b cluster length with
respect to sizes of macromolecule. We will demonstrate that MBE(3)-OSV-MP2
computation suffices to achieve a similar accuracy to what the direct OSV-MP2
energy and gradients can have with identical OSV cut-off $l_{\text{osv}}$. 

The selection and screening schemes of 2b and 3b clusters are essential for
lowering costs of expensive OSV-MP2 gradient terms. Here, the 2b and 3b
expansions are truncated based on the algorithmic metric between the OSVs
associated respectively with each LMO, which avoids caveats from real space
measurements. As the locality of LMOs and the compactness of OSVs facilitate an
exponential decay of OSV overlap matrix elements with the pair $ij$ separation,
the average square norm of the OSV overlap matrix indicates the pairwise
interaction strength between $ij$ OSV domains that constitute the 2b cluster
\begin{equation}
 s^\text{2b}_{ij} = \frac{\sum_{\bar\mu \bar\nu}\braket{\bar\mu_{i}|\bar\nu_{j}}^2}{\sqrt{n_i n_j}}
\label{eq:2bscreen}
\end{equation}
with $\braket{\bar\mu_{i}|\bar\nu_{j}}=\mathbf{Q}_i^\dagger\mathbf{Q}_j$ and
$n_i$ the total number of OSVs for the $i$-th LMO. Using the relation $n_i =
\sum_{\bar\mu\bar\nu} \braket{\bar\mu_{i}|\bar \nu_{i}}^2$, OSV orthonormality
and Cauchy-Schwarz inequality, there must be off-diagonal elements 
$0\le s^\text{2b}_{ij}\le 1$ and diagonal elements $s_{ii}=1$. 
Apparently, the magnitude of $s^\text{2b}_{ij}$ exhibits strong dependence on
the choice of kept OSVs, for instance, $s^\text{2b}_{ij} \to 1$ when the OSV set
becomes more complete.  This ensures that more strong 2b clusters enclosing
important pairwise interactions can be automatically identified and adaptively
selected, when it is necessary to employ extended OSVs due to tighter OSV
cut-off $l_{\text{osv}}$ or more delocalized nature of orbitals.  The selection of 3b
clusters which contain the united OSV sets for $ijk$ LMOs is based on the
mean of the pairwise metrics $s^\text{2b}_{ij}$, $s^\text{2b}_{ik}$ and
$s^\text{2b}_{jk}$ from the respective 2b clusters,
\begin{equation}
  s^\text{3b}_{ijk} = \frac{1}{3}\left (s^\text{2b}_{ij}+s^\text{2b}_{ik}+s^\text{2b}_{jk} \right).
\label{eq:3bscreen}
\end{equation}
Given a prescription $l_\text{2b}$ and $l_\text{3b}$ for choosing 2b and 3b
clusters, only those important 2b and 3b clusters with $s^\text{2b}_{ij}$ and
$s^\text{3b}_{ijk}$ exceeding their respective $l_\text{2b}$ and $l_\text{3b}$
values are kept for explicit OSV-MP2 energy and gradient computations.
Nevertheless, since the discarded weak 2b corrections  amount to still
considerable contributions, swift and accurate long-range 2b corrections are
implemented and will be presented in the ensuing section.  The MBE(3)-OSV-MP2
computation is therefore virtually controlled through a combination of three
simple parameters: $l_{\text{osv}}$, $l_\text{2b}$ and $l_\text{3b}$ for
selection of OSVs, 2b and 3b clusters, respectively. However, $l_\text{3b}$ must
be large enough, as compared to 2b clusters, to allow only 3b clusters of
sufficiently strong pairwise interactions.  Compared to canonical reference
results, we find that the MBE(3)-OSV-MP2 parameters by $l_{\text{osv}}=10^{-4}$,
$l_\text{2b}=10^{-2}$ and $l_\text{3b}=0.2$ yield correlation energies at
accuracy better than $99.95\%$ for small testing molecules and $99.7\%$ for
large molecules, and gradient RMSDs (root-mean-square deviation) below $10^{-4}$
au.

A major challenge of implementing MBE(3)-OSV-MP2 gradient theory is that it
incurs computations of many AO components on the full scale of macromolecule, if
separate MBE(3) gradients in eq~\ref{eq:mbe2} are carried out on a term-by-term
basis, for instance, by repeatedly evaluating AO-based gradients
eq~\ref{eq:ec_grad} and solving Z-vector eq~\ref{eq:zvec} for each
differentiation. It is essential to confer an implementation in which we can
perform a nonredundant set of small and rapid MBE(3) computations that are
unique to individual 2b and 3b clusters in the OSV basis, and evaluate these
AO-driven equations once and for all. The idea is to apply the above MBE(3)
partitioning and clustering to selected intermediates with major computational
costs, rather than to term-by-term energy gradients. These small pieces are then
collected to assemble the one-electron contributions $\mathbf{\check D}$ and
$\mathbf{\check D'}$, as well as the two-electron contribution $\mathbf{Y}_i$.

We can divide these intermediates into 1b-, 2b- and 3b-specific variables.
Apparently, LMOs $\mathbf{C}_i$, OSVs $\mathbf{Q}_i$, 3c2e AO integrals
$\mathbf{J}_i$ and derivative integrals $\mathbf{J}_i^{\lambda}$ are 1b-specific
and usually generated when computing 1b clusters; the OSV overlap
$\mathbf{S}_{(ij,ij)}$, the Fock $\mathbf{F}_{(ij,ij)}$, the OSV density matrix
$\mathbf{D}_{(ij,ij)}$ and two-electron relaxed amplitudes $\mathbf{X}_{ij}$ are
2b-specific, which are determined explicitly up to 2b clusters in terms of other
2b- and 3b-specific objects; the 3b-specific objects, which exhibit the
dependence on the explicit pairwise 2b interactions and meanwhile explicit extra
interactions correlating more LMOs beyond the pair, as seen in the OSV-MP2
residual in eq~\ref{eq:rij} and and its OSV response equations in \ref{eq:nij}.
We therefore applied MBE(3) scheme to the following 3b-specific objects, i.e.,
the  OSV amplitudes $\mathbf{T}_{(ij,ij)}$, the internal density matrix $D_{ij}$
and the intermediate $N_{ij}$ as described using MBE(3) expansion in
eqs~\ref{eq:tmbediag}--\ref{eq:nmbeij} analogous to energy, by which the pairwise
density matrices (e.g., $\mathbf{D}_{(ij,ij)}$, $\mathbf{D}^v$ and
$\mathbf{D}'^v$, $\mathbf{D}^o$ and $\mathbf{D}'^o$) and two-electron relaxed
amplitudes  $\mathbf{X}_{ij}$ can be then computed only at 2b level.  The
following MBE(3) expansion has been developed for the diagonal
\textit{collective} pair amplitudes $\mathbf{T}_{(ii,ii)}$,
\begin{equation}
\mathbf{T}_{(ii, ii)} = \mathbf{T}^i_{(ii,ii)}  + \sum_{k}\Delta\mathbf{T}^{i,k}_{(ii,ii)} 
                        + \sum_{k>l}\Delta\mathbf{T}^{i,k,l}_{(ii,ii)}, 
\label{eq:tmbediag}
\end{equation}
\begin{equation}
\Delta\mathbf{{T}}^{i,k}_{(ii,ii)}  = \mathbf{{T}}^{i,k}_{(ii,ii)}-\mathbf{T}^i_{(ii,ii)}, 
\end{equation}
\begin{equation}
\Delta\mathbf{T}^{i,k,l}_{(ii,ii)}  = \mathbf{{T}}^{i,k,l}_{(ii,ii)}
 -\Delta \mathbf{T}^{i,k}_{(ii,ii)} - \Delta \mathbf{T}^{i,l}_{(ii,ii)}
 - \mathbf{T}^i_{(ii,ii)},
\end{equation}
and the off-diagonal \textit{collective} pair amplitudes $\mathbf{T}_{(ij,ij)}$,
\begin{equation}
\mathbf{T}_{(ij,ij)} = \mathbf{T}^{i,j}_{(ij,ij)}+\sum_{k}\Delta\mathbf{T}^{i,j,k}_{(ij,ij)}, 
\label{eq:tmbeoff}
\end{equation}
\begin{equation}
\Delta\mathbf{T}^{i,j,k}_{(ij,ij)} = \mathbf{T}^{i,j,k}_{(ij,ij)}-\mathbf{T}^{i,j}_{(ij,ij)},
\end{equation}
where $\mathbf{T}^{i,j,k}_{(ij,ij)}$ and $\mathbf{T}^{i,k,l}_{(ii,ii)}$ are
solved independently from 3b clusters. For instance, the \textit{cluster}
amplitude $\mathbf{T}^{i,j,k}_{(ij,ij)}$ with the superscript $i,j,k$ is
obtained by solving the \textit{cluster} residual equation
$\mathbf{R}^{i,j,k}_{(ij,ij)}$ of the 3b cluster that encloses only $i,j,k$ LMOs
and associated OSVs.  Similarly, the MBE(3) expansions for $\mathbf{N}_{ij}$
follow
\begin{equation}
\mathbf{N}_{ii} = \mathbf{N}^i_{ii}  + \sum_{k}\Delta\mathbf{N}^{i,k}_{ii} 
                        + \sum_{k>l}\Delta\mathbf{N}^{i,k,l}_{ii}, 
\label{eq:nmbeii}
\end{equation}
\begin{equation}
\mathbf{N}_{ij} = \mathbf{N}^{i,j}_{ij}+\sum_{k}\Delta\mathbf{N}^{i,j,k}_{ij}
\label{eq:nmbeij}
\end{equation}
where $\mathbf{N}^{i,j,k}_{ij}$ is computed according to eq~\ref{eq:nij} taking
only $i,j,k$ LMOs for the $ij$ pair. The MBE(3) expansion is also similarly
carried out for $D_{ij}$.
The MBE(3) scheme facilitates massive parallel computations of these
small increments in eqs~\ref{eq:tmbediag}--\ref{eq:nmbeij} by distributing independent
tasks on many processes, which will be discussed in section~\ref{sec:parallel}.

\subsubsection{Correlation scheme for weak 2b clusters}

\begin{figure}
\centering
\includegraphics[scale=0.40]{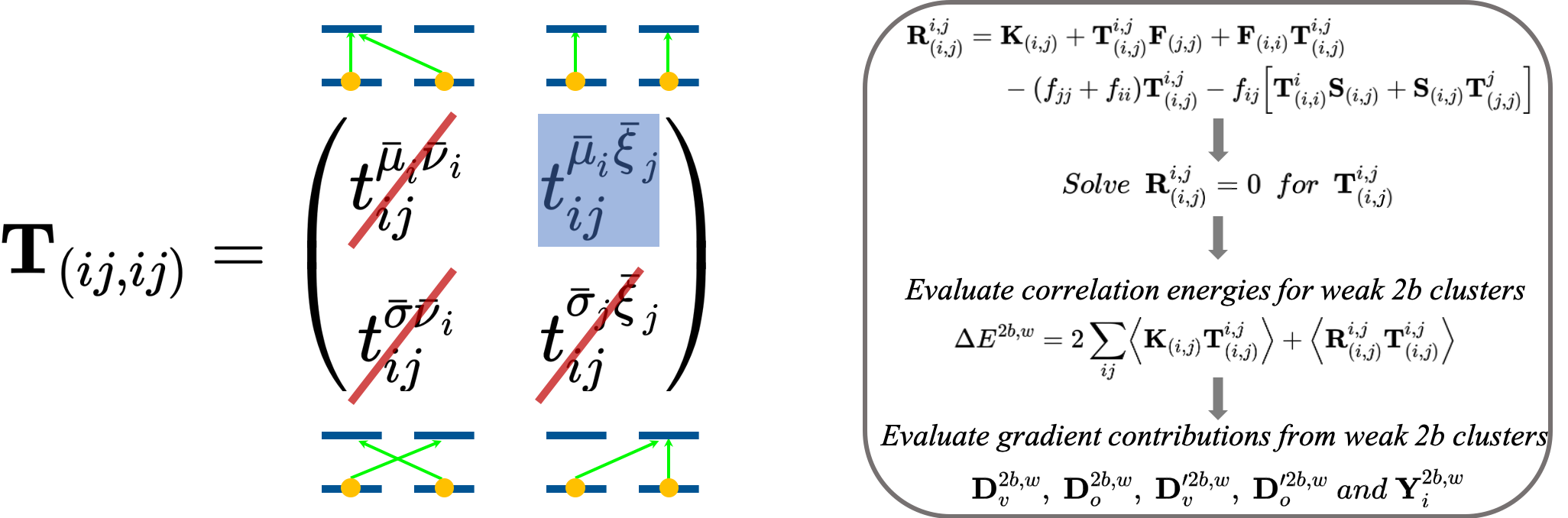} 
\caption{The excitation classes encoded in OSV-MP2 amplitudes (left) and the
one-block correlation scheme for treating weak 2b clusters (right).}
\label{fig:weak2b}
\end{figure}

Since the number of full 2b clusters increases quadratically with molecular
size, to convert the 2b computations into a practically tractable problem, we
choose only a subset of 2b clusters for rigorous OSV-MP2 computations, according
to the 2b screening scheme in eq~\ref{eq:2bscreen}. A large number of weak 2b
clusters, if all simply omitted, would nevertheless produce a substantial amount
of aggregate errors to both correlation energies and gradients, which presents a
major obstacle for realizing reliable MBE(3)-OSV-MP2 algorithm on large
molecules. However, the convergence of the long-range correlation existing in a
weak 2b cluster is asymptotically dominated by direct dispersion rather than
charge transfer and exchange correlation components. This family of correlation
contributions is typically described by the four subblocks of the cluster
amplitudes $\mathbf{T}_{(ij,ij)}$, with distinct excitation classes depicted
in Figure~\ref{fig:weak2b}.
The dispersion possesses $\{i\to \bar\mu_i,j\to \bar\xi_j\}$
double excitations genuinely represented by $\mathbf{T}_{(ij,ij)}$ upper-right
block, while the diagonal and the lower-left blocks are responsible for the
charge transfer $\{i\to \bar\mu_i, j\to \bar\nu_i \}$ and
exchange $\{i\to \bar\sigma_j,j\to\bar\nu_i\}$ excitations,
respectively.  

For the above reasons, we resolve the weak 2b cluster problems by projecting the
upper-right block out of the full 2b residual equations, which leads to the
one-block algorithm in which the one-block residual equations
$\mathbf{R}^{i,j}_{(i,j)}$ are solved for one-block 2b amplitudes
$\mathbf{T}^{i,j}_{(i,j)}$, aiming for swiftly capturing direct dispersion. 
For the gradient contributions from weak 2b clusters, the relevant intermediates
of one-block structure of eq~\ref{eq:dv} are needed. For instance,
\begin{equation}
  \mathbf{D}^v_{\text{2b},w} = \sum_{ij} \mathbf{Q}_i \mathbf{D}_{(i,j)} \mathbf{Q}_j^\dagger
                        + \sum_{ij} \mathbf{T}_{ii} \mathbf{X}_{ij}^{\text{2b},w}
\end{equation}
with one-block overlap- and energy-weighted  matrices in the OSV basis,
\begin{equation}
\mathbf{D}_{(i,j)_w} = \mathbf{T}^{i,j}_{(i,j)}\mathbf{S}_{(j,i)}\mathbf{T}^{i,j}_{(i,j)} +
 \mathbf{T}^{\dagger i,j}_{(i,j)}\mathbf{S}_{(j,i)}\mathbf{T}^{\dagger i,j}_{(i,j)},
\quad \mathbf{D}'_{(i,j)_w} = \mathbf{T}^{i,j}_{(i,j)}\mathbf{F}_{(j,i)}\mathbf{T}^{i,j}_{(i,j)} +
 \mathbf{T}^{\dagger i,j}_{(i,j)}\mathbf{F}_{(j,i)}\mathbf{T}^{\dagger i,j}_{(i,j)}
\end{equation}
and the one-block analogue of the residual relaxation is accounted by
$\mathbf{X}_{ij}^{\text{2b},w}$.
We find that the one-electron part of the residual response in eq~\ref{eq:nij}
makes virtually indiscernible contributions to the total gradient. The
insignificance of the one-electron contribution to the residual relaxation
results from the small overlap matrix between OSVs residing in the proximity of
the remote LMOs that constitute weak 2b clusters. This is demonstrated, for
instance, to nonactin molecule for which when 8214 out of 11026 LMO pairs are
treated as weak 2b clusters, the
OSV-MP2/$l_{\text{osv}}=10^{-4},l_{\text{2b}}=10^{-2}$/def2-tzvp RMSD is only
$1.2\times 10^{-5}$ between the gradients with and without one-electron residual
relaxation, and the maximum deviation only $8.3 \times 10^{-5}$.  For reasons of
CPU, memory and I/O efficiency, the following approximation is implemented for
upper and lower blocks of $\mathbf{N}_{ij}^{\text{2b},w}$, respectively,
\begin{equation}
\left(\mathbf{N}_{ij}^{\text{2b},w}\right)^\text{up} \approx
\mathbf{T}^{i,j}_{(i,j)}\mathbf{K}_{(j,i)},
\quad \left(\mathbf{N}_{ij}^{\text{2b},w}\right)^\text{low}\approx 0
\label{eq:nij2b}
\end{equation}
The projected one-block correlation scheme leads to much reduced complexities of
computing residual and gradient intermediates belonging to weak 2b clusters,
formally with only about $1/16$ and $1/8$ of the costs for computing each strong 2b
cluster, respectively, which is therefore comparatively negligible.

Moreover, extremely remote 2b clusters that are even weaker are all discarded
when the pairwise interaction strength $s_{ij}^\text{2b}$ is below $10^{-7}$.
This can sometimes (but not always) discard a large amount of insignificant 2b
clusters, for instances, there are 13019 strong, 144247 weak and 131914
discarded 2b clusters for (H$_2$O)$_{190}$,  5970 strong, 13870 weak and 78950
discarded 2b clusters for (Gly)$_{40}$, but for C$_{60}$@catcher molecule, there
are 7190 strong, 25179 weak and only 16 discarded 2b clusters.

\subsubsection{\label{sec:spar}Sparsity for two-electron integrals and OSV relaxation}


For efficient evaluations of energy and gradient, we implemented an integral
algorithm for performing the parallel
half-transformation $(\alpha\beta|A)\rightarrow(i\alpha|A)$ in multiple tasks
according to AO shell pairs of $(\alpha\beta|A)$ that are prescreened using
Cauchy-Schwarz relation $\left|(\alpha\beta|A)\right| \leq
\left\Vert{(\alpha\beta|\alpha\beta)}\right\Vert \cdot
\left\Vert{(A|A)}\right\Vert$.  
However, the next fitting step $J_{i,A\alpha}=\sum_B (i\alpha|B){V}^{-1/2}_{AB}$
needed for computing the two-electron response intermediate $\mathbf{Y}_i$ in
eq~\ref{eq:ec2eY} requires the transformation with the Coulomb metric
${V}_{AB}=(A\vert B)$ with high operational cost $ONN^2_{aux}$ scaling up
quadratically with the size of auxiliary functions for large molecules.
In the context of local correlation methods, this problem can be circumvented
for exchange integral transformation by selecting a union of local fitting and
atomic orbital domains associated with the occupied $ij$ pairs, i.e., $A,B \in
[ij]_\text{fit}$ and $\alpha\in [ij]_\text{AO}$ that help reduce the
scaling, known as local density fitting\cite{werner2003fast,werner2015scalable}.
Therefore both Cholesky decomposition of the metric $\mathbf{V}(ij)$ and fitting
steps must be carried out for each $ij$ pair. This certainly creates costly
overheads before proceeding to the next AO-to-OSV half-transformation 
$J_{i,A\alpha}\rightarrow J_{i,A\mu_j}$ which is sufficiently fast owing to the
short OSV and local auxiliary dimensions.  Moreover, the local fitting scheme is
not practical to the fitting of derivative integrals for energy gradients since
the half-transformation
$(\alpha\beta|A)^{(\lambda)}\rightarrow(i\alpha|A)^{(\lambda)}$, the
pair-specific Cholesky decomposition and fitting steps must be avoided for all
$3N$ nuclear positions. For consistent fittings of both integrals and the
corresponding derivatives, we have developed a sparse fitting strategy in which
the sparsity of $(i\alpha|A)$ and $(\alpha\beta|A)^{(\lambda)}$ are directly
exploited to reduce the amount of auxiliary functions that participate in
fitting $\mathbf{J}_i$ and $\mathbf{J}_{\alpha}^{(\lambda)}$, respectively.

For an occupied LMO $i$ assigned to each parallel process, only those auxiliary
functions $A'$ making important contributions to the fitting step are kept
according to the sum of square $(i\alpha|A')$ that must be greater than a
prescribed orbital-specific sparsity threshold $l_\text{ofit}$,
\begin{equation}
  \sum_{\alpha}(i\alpha|A')^{2} > l_\text{ofit}~\text{where }A'\in
[i]_\text{ofit}.\label{eq:lfit}
\end{equation}
Computing the sparsity of eq~\ref{eq:lfit} adds negligible costs due to the
small vector size in each parallel batch $[i]$, and thus the full sparsity of
$(i\alpha|A')$ can be efficiently utilized for fitting $\mathbf{J}_i$.  By
construction, the sparse fitting domain $[i]_\text{ofit}$ is orbital-specific
and only necessitates the mergence of
$[ij]_\text{ofit}=[i]_\text{ofit}\cup[j]_\text{ofit}$ for pair $ij$ when
assembling $\mathbf{J}_i^\dagger\mathbf{J}_j$ for energy and
$\mathbf{J}_i^\dagger\mathbf{Y}_j$ for gradient. Our numerical experiments show
that the merged sparse fitting domain is only moderately larger than the
pair-specific fitting domain of local fitting method with comparable accuracy of
energy and gradient.  The orbital-specific sparse fitting scheme also
significantly accelerates the computation of exchange integrals needed for
Z-vector, i.e., the last two exchange potentials in $A_{ai,bj}$ of
eq~\ref{eq:zveca}. 

The computation of $\sum_{bj}(aj\vert bi)Z_{bj}$ is straightforward by a single
$\mathbf{J}_i$ fitting based on the auxiliary selection eq~\ref{eq:lfit}.
However, in order to avoid double fittings for $(ab\vert ij)$ in the presence
of asymmetric 3c2e half-integrals, its transformation with Z vector is computed
as follows,
\begin{equation}
  \sum_{bj}(ab|ij)Z_{bj} = \sum_{bj, A''}(ab|A'')Z_{bj}{J''}_{i,A''j}
\end{equation}
and $\mathbf{J''}_i$ is obtained by solving the linear equation
\begin{equation}
\sum_{B''}V_{A''B''}J''_{i,B''j} = (ij \vert A'').
\end{equation}
Here the auxiliary functions $A''$ are selected according to the predefined
block sparsity $l_\text{bfit}$
\begin{equation}
  \max_{A''\in [i]_\text{bfit}}{\sum_{\alpha}(i\alpha|A'')^{2}}>l_\text{bfit}.
   \label{eq:lbfit}
\end{equation}

Numerical tests for Nonactin/def2-tzvp (C$_{40}$H$_{64}$O$_{12}$, 116 atoms)
show that the gradient accuracy is hardly affected by loosening the block sparsity
$l_\text{bfit}$. As shown in Table~\ref{tab:sparse}, given $l_\text{ofit}=10^{-6}$, the
maximum absolute error and RMSD in analytical gradient deviations are almost
unchanged from $l_\text{bfit}=10^{-4}$ to $l_\text{bfit}=10^{-2}$, compared to
results without using sparsity, and the looser $l_\text{bfit}=10^{-2}$ greatly
improves the scaling behaviour of exchange integral transformation in Z-vector
computation. Overall, we find that $l_\text{ofit}=10^{-6}$ and
$l_\text{bfit}=10^{-2}$ make reasonable sparsity thresholds and are applied to
integrals for OSV generation, exchange integral transformation, derivative
integrals and Z-vector solution, for which an average two-fold speedup was
observed.
\begin{table}[H]
  \caption{\label{tab:sparse}Comparison of fitting sizes, elapsed time (sec),
the correlation energy (Percentage, \%) and gradient (MAXD/RMSD, au) accuracy
with respect to the orbital-specific sparse and block fitting thresholds for
Nonactin molecule (C$_{40}$H$_{64}$O$_{12}$) using def2-tzvp basis set. All
parallel computations were carried out on 24 CPUs.}
  \begin{tabular}{ccccccc}
  \hline
  \hline
$l_{\text{ofit}}$/$l_{\text{bfit}}$&0/0&$10^{-6}/10^{-4}$&$10^{-6}/10^{-3}$&$10^{-6}/10^{-2}$&$10^{-5}/10^{-5}$&$10^{-4}/10^{-4}$\\
\hline												
RHF $N_{\text{ofit}}$ per LMO	&	5076	&1674	&	1674	&	1674	&	1009	&	522	\\
RHF $N_{\text{bfit}}$ per LMO	&	5076	&2992	&	1467	&	575	&	2231	&	1467	\\
MP2 $N_{\text{ofit}}$ per pair	&	4912	&1686	&	1686	&	1686	&	1148	&	673	\\
$t_{\mathbf{T}_{kk}}$         	&	8.2	&4.9	&	5.1	&	4.9	&	3.2	&	2.0	\\
$t_{\mathbf{K}_{(ij,ij)}}$ 	&	19.2	&12.4	&	12.7	&	12.2	&	9.4	&	6.8	\\
$t_{\mathbf{K}^{\{\lambda\}}_{(ij,ij)}}$ &36.6	&	22.0	&	22.4	&	21.8	&	15.2	&9.8	\\
$t_\mathbf{K}$           	&	210.0	&	149.1	&	108.7	&	80.7	&	126.5	&	104.3	\\
Percentage 			&	100	&	99.99	&	99.99	&	99.99	&	99.94	&	99.56	\\
MAXD ($10^{-4}$)    		&	0.0  	&	5.1  	&	5.1  	&	5.1  	&	14  	&	97    \\
RMSD ($10^{-4}$)    		&	0.0  	&	1.0  	&	1.0  	&	1.0  	&	3.2 	&	20    \\
\hline
\hline
  \end{tabular}
\end{table}

The OSV derivative relaxations of two-electron integrals and OSV-MP2 residuals
occur (via the intermediate $\mathbf{N}_{ij}$ in eq~\ref{eq:nij}) between the
kept and discarded OSV subspaces, which has unfavorable costs for large
molecules due to a large number of discarded OSVs. The important OSV relaxation
vectors making most contributions to OSV-MP2 gradients can be selected based on
the intrinsic sparsity amongst the discarded OSV vectors $\mathbf{Q}'_i$. Here
we adopt an interpolative decomposition (ID)
estimate~\cite{liberty2007randomized} to rapidly generate approximate OSVs
(ID-OSVs) from numerically low-rank MP2 diagonal amplitudes $\mathbf{T}_{kk}$,
prescribed with a rank cutoff $l_{\text{cposv}}$ for automatically identifying an
important subset of each $\mathbf{Q}'_i$. This particularly reduces the cost of
OSVs generation from original $N^4$ for exact OSVs to $cN^2$ for ID-OSVs on all
occupied MOs, with the prefactor $c$ determined by $\mathbf{T}_{kk}$ rank
according to $l_{\text{cposv}}$. For instance, for Nonactin molecule using
def2-tzvp basis (Table~S1), the ID-OSV/$l_{\text{osv}}=10^{-4}$
generation with $l_{\text{cposv}}=10^{-4}$ gains a nearly seven-fold speedup
compared to exact OSVs/$l_{\text{osv}}=10^{-4}$, yielding only a minor loss of
correlation energy by $7.3\times 10^{-6}$ au. However, a tight
$l_{\text{cposv}}=10^{-10}$ is desired for very accurate analytical gradients
which typically halves $\mathbf{Q}'_i$ vector, leading to only  gradient RMSD of
$2.6\times 10^{-5}$ au. For most applications,
$l_{\text{cposv}}=10^{-6}$--$10^{-7}$ is a normal choice which guarantees reasonably
accurate gradients around $10^{-4}$ au and fast OSV generation and relaxation.
When the extremely tight $l_{\text{cposv}}=10^{-10}$ is needed for targeting
highly accurate gradients, e.g., RMSD $\sim10^{-5}$ au, which is however very
rare for large molecules, a direct selection scheme for $\mathbf{Q}'_i$ is
preferred based on exact OSVs since the ID convergence of low-rank
$\mathbf{T}_{kk}$ becomes slow and the computational saving is lost
unfortunately.  

\subsubsection{\label{sec:parallel}Parallel implementation}

The MBE(3)-OSV-MP2 necessitates parallel computations of all energy and gradient
corrections up to the third-order. Our parallelism and implementation details
are presented in Appendix. While it is always the perfection and
sophistication of runtime balance between memory usage, disk storage, data
communication and costs in duplicated computing tasks that achieves high-level
scalable parallelization, we harness the parallel efficiency by primarily aiming
for accessibility and affordability of remote/local (shared) memories amenable
to large molecules. In the current parallel implementation for MBE(3)-OSV-MP2, a
multi-node parallelism is built in Message Passing Interface (MPI) standard of
version 3~in which low-latency one-sided intra- and inter-node communications
within the memory region accessible to all remote processes were exploited.
This is significantly faster with lower data communication latency than
traditional point-to-point MPI communication by reducing individual memory copy
operations and synchronizations occurring in the communication from/to each
remote process using passive targets.  Here, we assume that broad bandwidth
inter-node connection (such as Infiniband) is nowadays readily available for
high performance computation of large molecules, whereby we do not distinguish
intra- and inter-node processes in the current implementation.  To further
reduce the synchronization time, the MBE(3)-OSV-MP2 amplitude clusters are
sorted according to the total OSV sizes and then distributed to all processes as
evenly as possible, so that the computational tasks assigned to each process are
as close as possible.

The data parallelism is based on the hybrid remote memory access (RMA) and
shared memory (SHM) mechanisms. RMA is enabled by constructing global array as
the partitioned Global Address Space (pGAS) accessible by processes of global
rank on multiple nodes. The pGAS is expanded incrementally with the number of
nodes for sharing and transferring increasingly large intermediates with sizes of
molecule. However, since  routine computations for large molecules are normally
performed on a limited number of nodes, it is unrealistic to enable a huge pGAS
for all distributed data objects. Thus, only the tensorial quantities in OSV
basis can be accessed globally, including $\mathbf{Q}_i$ vectors,
$\mathbf{S}_{(ij,ij)}$, $\mathbf{F}_{(ij,ij)}$, $\mathbf{X}_{(ij,ij)}$, integrals
$\mathbf{K}_{(ij,ij)}$ and their OSV geometric relaxation, and the OSV amplitudes
$\mathbf{T}_{(ij,ij)}$. Additionally, the integral-incore implementation for
medium size molecules also places the  half-transformed MO 3c2e integrals, MP2
diagonal amplitudes $\mathbf{T}_{kk}$, the discarded OSV vectors $\mathbf{Q}'_i$
and residual response $\mathbf{Y}_i$ in pGAS, and otherwise they are stored on
disk in the integral-direct algorithm. Finally, an SHM window is allocated to the
root process for matrices of lower dimension than RMA tensors, e.g., Coulomb
matrix $V_{AB}$, OSV-MP2 density matrices, $J_{\alpha i}$ and $K_{\alpha i}$ for
Z-vector potentials, which can be accessed by other processes within the node.
As such, the root process is conveniently utilized for matrix update and
accumulation as needed, by harvesting data from other processes within the node.

In a typical OSV-MP2 gradient computation, the major time is spent in the
evaluation of two-electron contributions to the gradient, needing unique terms
of $\sum_i \mathbf{P}_{v}\mathbf{Y}_{i}^{\dagger}\mathbf{J}_{i}$,
$\braket{\mathbf{P}_{v} \mathbf{Y}_{i}^{\dagger} \mathbf{J}_{p}}$ and
$\sum_i\braket{\mathbf{P}_{v}\mathbf{Y}_{i}^{\dagger}\mathbf{J}_{i}^{(\lambda)}}$,
which requires the 3c2e RI AO integrals $\mathbf{J}_{p}$, the AO derivatives
$\mathbf{J}_{i}^{(\lambda)}$ and the intermediate $\mathbf{Y}_i$
(eq~\ref{eq:ec2eY}). In our previous serial implementation, they were computed
explicitly and stored in memory or on disk, which was convenient for
small-to-medium sizes of molecule. Consider a large water cluster
(H$_2$O)$_{190}$, the large intermediates $\mathbf{Y}_i$ of the dimension
$OVN_{aux}$ for all LMOs, which are about $800$ Gb for (H$_2$O)$_{190}$/cc-pvdz
basis and $2400$ Gb for (H$_2$O)$_{190}$/cc-pVTZ in size, result in rather
unfavorable storage and I/O overheads which should be avoided. Repeated
computations of $\mathbf{Y}_i$ for each gradient contribution are not desirable
due to high cost of ${\cal{O}}(O^2N_{osv}VN_{aux})$, even asymptotically with
$c{\cal{O}}(N^{3})$ with screened LMO pairs. However, $\mathbf{Y}_i$ can be
easily vectorized with respect to multi-node batches of the auxiliary shells,
with each task of $\mathbf{Y}_i(A)$ short enough for transformations as follows, 
\begin{equation}
 y_{i\alpha} = \sum_{A}^\text{tasks}\sum_\gamma Y_{i,A\gamma} J_{\alpha,A \gamma},\quad
 y'_{\alpha\beta} = \sum_{A}^\text{tasks}\sum_i Y_{i,A \alpha} J _{i, A \beta}
\label{eq:yint}
\end{equation}
where one-sided accumulations of $\mathbf{y}$ and $\mathbf{y'}$
from different processes are carried out. For each process where a small number
of $A$ auxiliary functions ($n_A$) reside, these transformations in
eq~\ref{eq:yint} incur $2n_AON^2$ operations and small $OV+N^2$ storage holding
$\mathbf{y}$ and $\mathbf{y'}$.  The two-electron contributions are then
collected,
\begin{eqnarray}
  \sum_i \mathbf{P}_{v}\mathbf{Y}_{i}^{\dagger}\mathbf{J}_{i}&=& \mathbf{P}^v \mathbf{y'},\\ 
\braket{\mathbf{P}_{v} \mathbf{Y}_{i}^{\dagger} \mathbf{J}_p} &=& \braket{\mathbf{P}^v\mathbf{yC}_p}, \\
\sum_i\braket{\mathbf{P}_{v}\mathbf{Y}_{i}^{\dagger}\mathbf{J}_{i}^{(\lambda)}} &=&
  \braket{\sum_{i\alpha}\left[ C_{\alpha i}\mathbf{P}^v\mathbf{Y}_i^\dagger \right]\mathbf{J}_\alpha^{(\lambda)}}.
\label{eq:jlambda}
\end{eqnarray}
Eq~\ref{eq:jlambda} formally costs $2n_AON^2+3n_An_\text{atom}N^2$ and forms the
most expensive step in eqs~\ref{eq:yint}--\ref{eq:jlambda}. However, the scaling
of $3n_An_\text{atom}N^2$ due to trace operation can be further lowered by
exploring the sparsity fitting with derivative integrals
$\mathbf{J}_\alpha^{(\lambda)}$ in a straightforward manner.

\section{NUMERICAL ASSESSMENTS}

\subsection{\label{sec:mbeerror}MBE$(3)$-OSV-MP2 Cluster Errors}

\subsubsection{Energy, gradient and structure}

\begin{figure}[H]
  \centering
  \begin{minipage}[b]{8.5cm}
   \includegraphics[width=8cm,trim={0.2cm 0 1.5cm 0},clip]{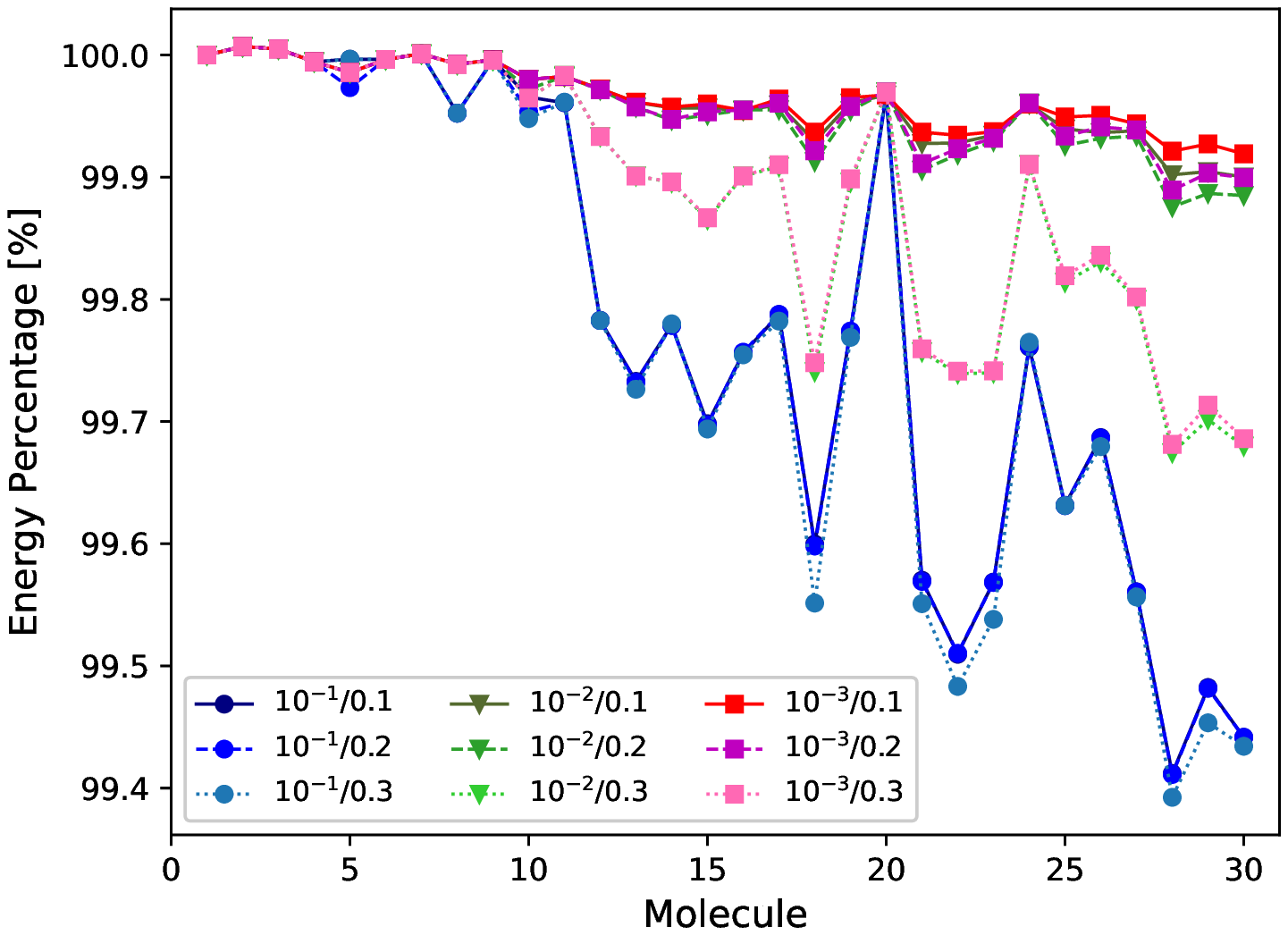}
   \caption*{(a)}
  \end{minipage}%
  \begin{minipage}[b]{8cm}
   \includegraphics[width=8cm,trim={0.2cm 0 1.5cm 0},clip]{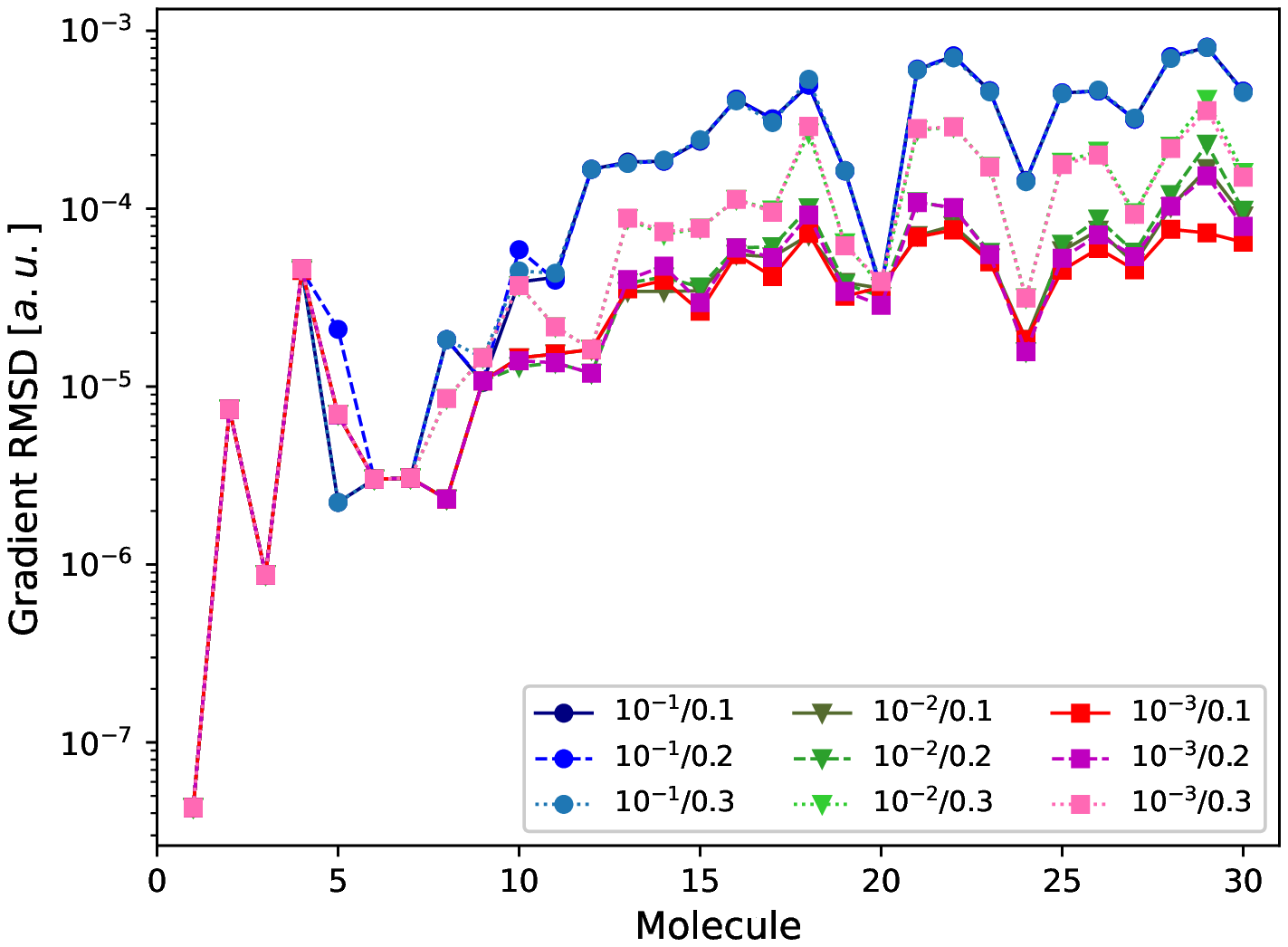}
   \caption*{(b)}
  \end{minipage}

  \caption{Comparisons of the percentages of correlation energy (a) and the
RMSDs (root mean square deviations) of gradient (b) between the MBE(3)-OSV-MP2
($l_\textrm{osv}=10^{-4}$) and canonical RI-MP2 with respect to the 2b and 3b
cluster selections ($l_\textrm{2b}/l_\textrm{3b}$). The basis set def2-tzvp was
used for all computations.}
  \label{fig:thres_sel}
\end{figure}

Efficient computations of MBE(3)-OSV-MP2 energy and gradient necessitate a
reasonable selection of 2b and 3b clusters according to the cluster criteria of
eqs~\ref{eq:2bscreen} and \ref{eq:3bscreen} which also depend on the choice
of OSVs.  The normal OSV choice $l_\textrm{osv}=10^{-4}$ was shown to yield
typically at least 99.9\% MP2 correlation energy and $<10^{-4}$ au gradient
errors for Baker test molecules~\cite{baker1993techniques} of different sizes
and bonding types in our previous work\cite{zhou2019complete}. The convergence
of correlation energies and gradients RMSDs for MBE(3)-OSV-MP2 is again assessed
for these molecules with respect to 2b ($l_\textrm{2b}$) and 3b
($l_\textrm{3b}$) cluster selections.  As presented in
Figure~\ref{fig:thres_sel}, the RI-MP2 reference results of small molecules
containing up to 10 atoms are well reproduced within an energy gain better than
$99.9$\% and a gradient RMSD below $10^{-4}$ for all 2b and 3b cluster
selections. For larger molecules with more than 10 atoms, the loose 2b
($l_\textrm{2b}=10^{-1}$) and 3b ($l_\textrm{3b}=0.3$) selections lead to
greater errors in both energies ($99.8$--$99.4$\%) and gradients (a few
$10^{-4}$), and including more 3b clusters alone does not necessarily improve
the numerical accuracy, since the loss of many important 2b clusters at the
level of $l_\textrm{2b}=10^{-1}$ prevents the long-range orbital pairs from
entering 3b clusters, according to the numbers of 2b and 3b clusters shown in
Figure~S1.  A normal 2b/3b selection based on the combination of
$l_{\text{2b}}=10^{-2}$ and $l_{\text{3b}}=0.2$ yields much improved accuracy of
$>99.85\%$ energy percentages and $<3\times 10^{-4}$ gradient RMSDs for all
testing molecules, which are comparable to normal OSV-MP2 results.
This suggests that
$l_\textrm{osv}/l_{\text{2b}}/l_{\text{3b}}=10^{-4}/10^{-2}/0.2$ make reasonable
criteria for OSV and cluster selections and are thus used for the remaining
computations, unless otherwise noted.

\begin{figure}[H]
\begin{minipage}[b]{4cm}
\centering
  \includegraphics[width=4cm]{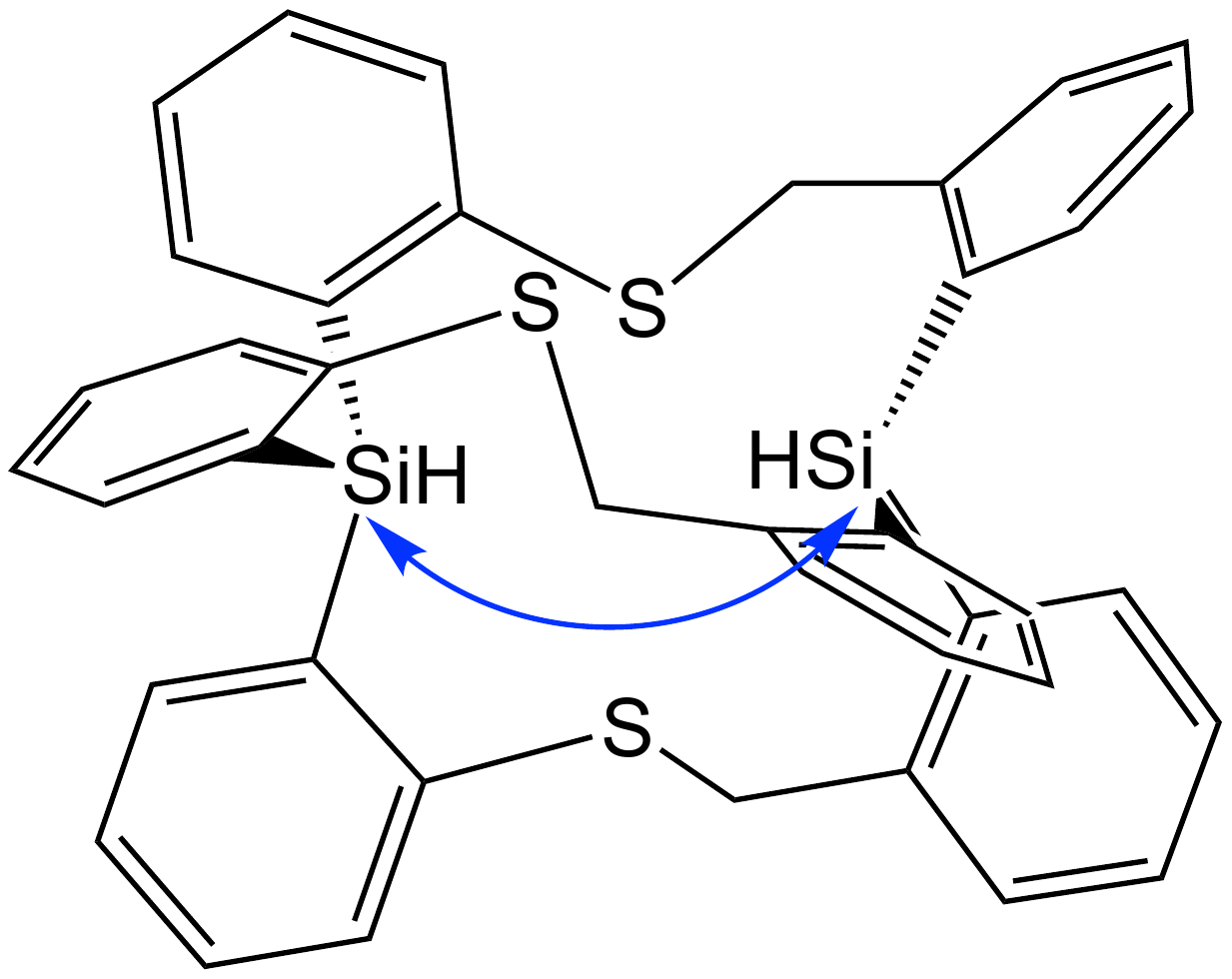}
   \caption*{BHS}
\end{minipage}
\begin{minipage}[b]{4cm}
\centering
  \includegraphics[width=4cm]{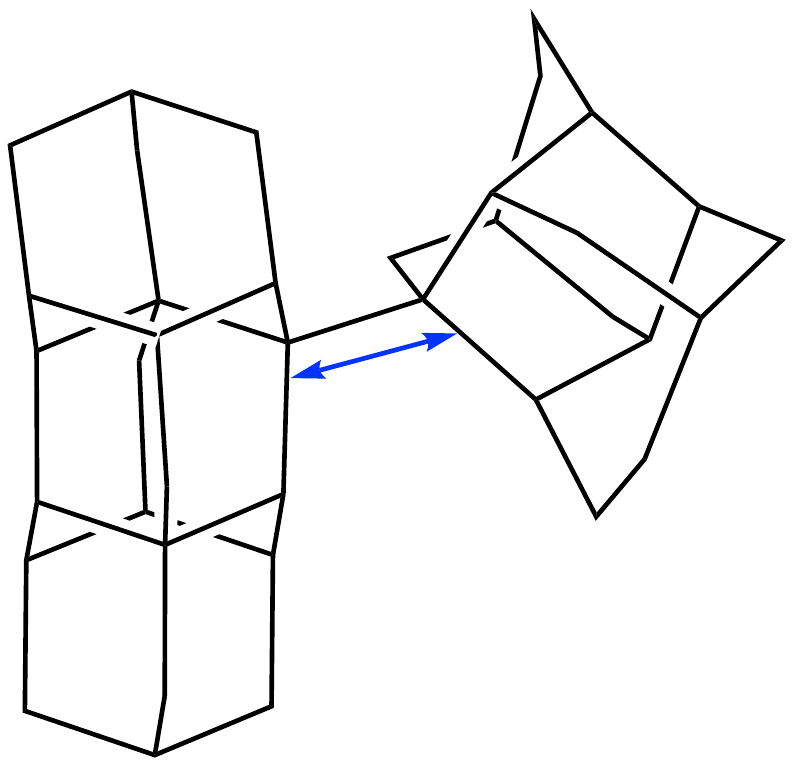}
  \caption*{DIAD}
\end{minipage}
\begin{minipage}[b]{4cm}
\centering
  \includegraphics[width=4cm]{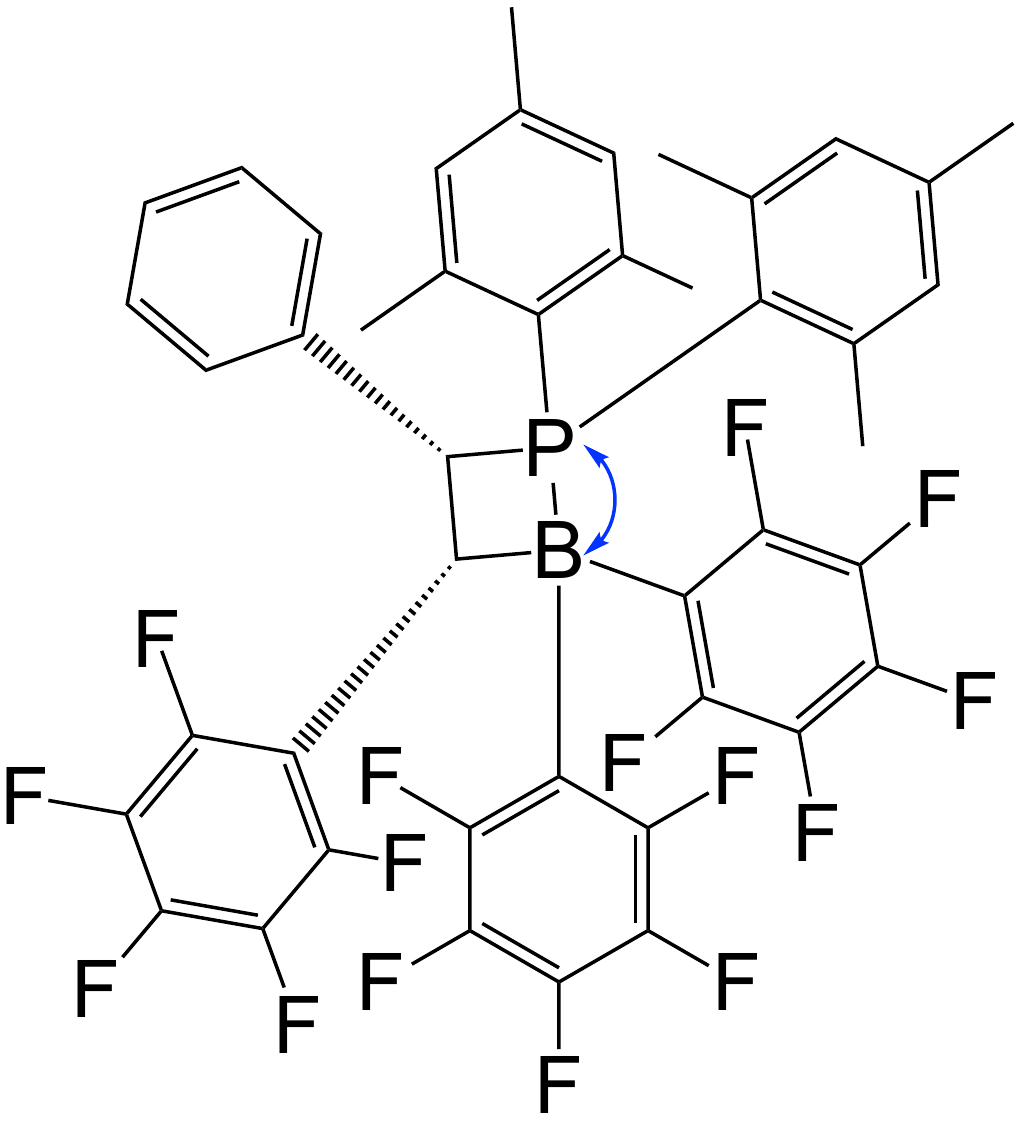}
   \caption*{FLP}
\end{minipage}
\begin{minipage}[b]{4cm}
\centering
  \includegraphics[width=4cm]{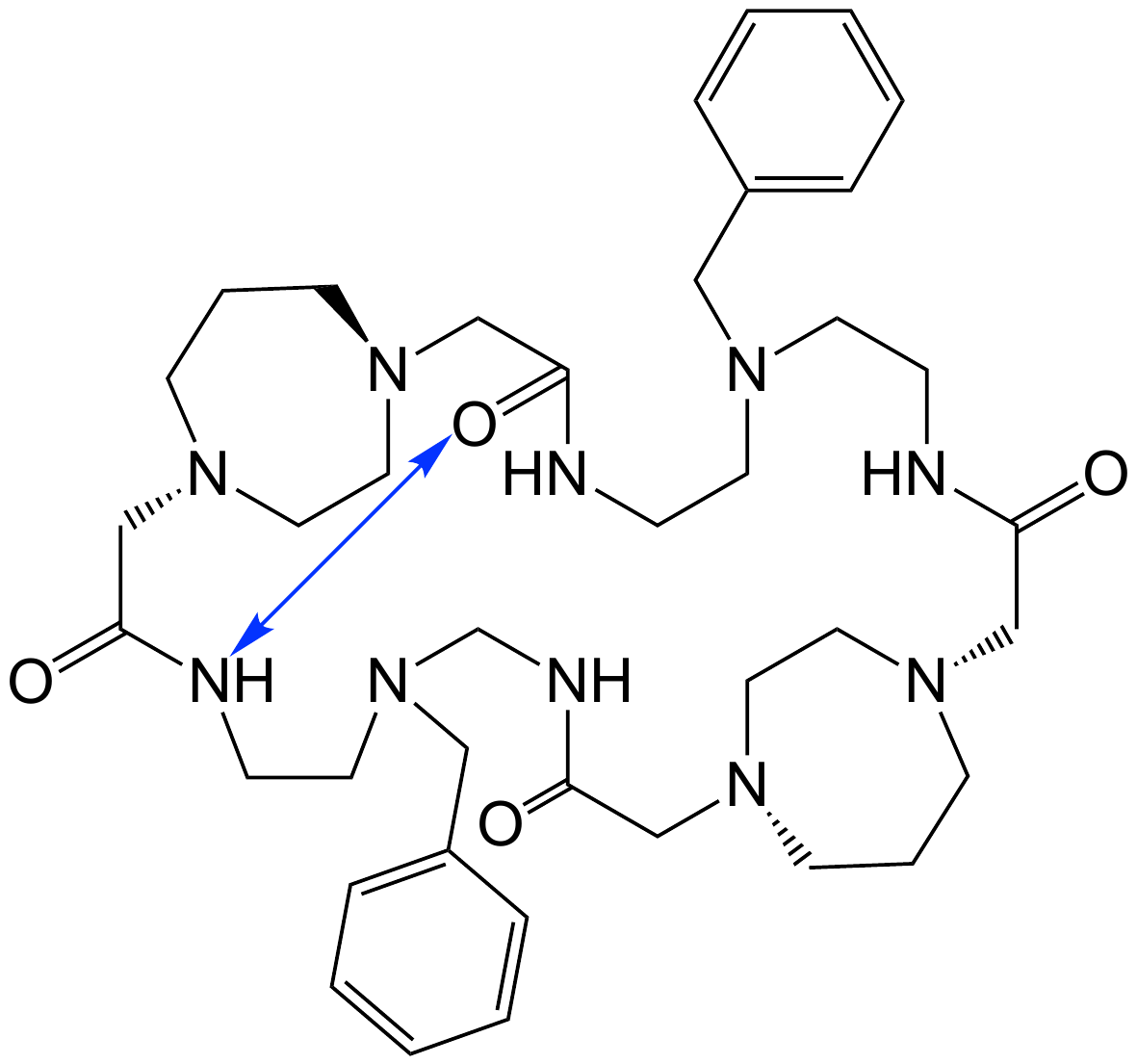}
  \caption*{YIVNOG}
\end{minipage}
  \caption{Medium size molecules for assessing MBE(3)-OSV-MP2 geometries for
which the canonical RI-MP2 reference structures can be computed.}
  \label{fig:f_mol_opt}
\end{figure} 

The MBE(3)-OSV-MP2 structures of several representative medium size molecules
containing different connectivities from second and third row elements (BHS,
FLP, DIAD and YIVNOG in Figure~\ref{fig:f_mol_opt}) are compared to RI-MP2
reference geometries. The deviations of selected interatomic distances are given
in Table~\ref{tab:t_bond_medium}. The maximal relative deviations from the
RI-MP2 interatomic distance are about 0.4\%, 0.3\%, 0.09\% and 0.4\% for BHS,
DIAD, FLP and YIVNOG, respectively, with the magnitudes varying between 0.1 and 1.7
pm.  The MBE(3)-OSV-MP2 accuracy is similar to that of normal OSV-MP2 using the same
$l_\textrm{osv}=10^{-4}$. The largest errors take place to BHS Si-Si distance
(1.7 pm) of C$_1$ symmetry and YIVNOG O-N distance (1.3 pm), both between non-bonded
atoms residing remotely on the periphery of the cavity. The MBE(3)-OSV-MP2 errors
for bonded atoms are however as small as about 0.5 pm for DIAD C-C and 0.1 pm
for FLP P-B bond. Overall, the MBE(3)-OSV-MP2 optimized structures
are sufficiently accurate compared to RI-MP2 benchmarks, and can be obtained by
terminating the MBE(3) expansion on a small amount of important 2b and 3b
clusters (Table~\ref{tab:t_bond_medium}). The improvements of bond lengths for
BHS and YIVNOG are very limited by including more 3b clusters using
$l_\text{3b}=0.1$, for which the numbers of 3b clusters are however considerably
increased from 6112 and 8560 to 9886 and 13852, respectively.

\begin{table}[H]
  \caption{\label{tab:t_bond_medium}Comparisons of correlation energy
percentages ($|\delta E_c|$) and optimized interatomic distance deviations
($|\delta d|$) between MBE(3)-OSV-MP2, normal OSV-MP2 and canonical RI-MP2 using
def2-tzvp basis and frozen core approximation. The structure convergence meets
three criteria: the energy change ($\le 10^{-6}$ au), the gradient RMS (root
mean square, $\le 3\times 10^{-4}$ au) and the maximum gradient ($\le 4.5\times
10^{-4}$ au). RI-MP2 structures were obtained using the ORCA program
package\cite{neese2018software} with RIJK integrals.
}
  \begin{tabular}{ccccc}
    \hline											
    \hline
molecules       	&BHS (Si-Si)	&	DIAD (C-C)&	FLP (P-B)&	YIVNOG (O-N)	\\
\hline											
$N_\text{atom}~^a$		& 76      	&	82	&	88	& 116	\\
$N_\text{orb}~^b$		& 1586		&	1392	&	2059	&2046	\\
$N_\text{aux}~^c$		& 4009		&	3426	&	5078	&5034	\\
$N_{2b}~^d$ 		& 5671		&	4465 	& 	12403 	& 10878 \\  
$N_{3b}~^e$ 		& 198485 	& 	138415 	& 	644956 	& 529396\\  [0.5em]
RI-MP2 reference   	&	441.4	&	168.2	&	209.6	& 307.8 \\[0.5 em]
\multicolumn{5}{c}{OSV-MP2 $l_\mathrm{osv}=10^{-4}$}\\
$|\delta E_c|$ (\%)	&	99.79	&	99.64	&	99.78	&	99.77 \\
$|\delta d|$ (pm) 		&	1.3	&	0.4 	&	0.2 	&	1.0   \\[0.5 em]
\multicolumn{5}{c}{MBE(3)-OSV-MP2 $l_\mathrm{osv}=10^{-4},l_\mathrm{2b}=10^{-2},l_\mathrm{3b}=0.2$}\\
$\tilde{N}_{2b}~^f$	& 1772	        &	2003	&	3213	&	2670	\\
$\tilde{N}_{3b}~^g$	& 6112	        &	7065	&	10785	&	8560	\\
$|\delta E_c|$ (\%)      	& 99.78	        &	99.65	&	99.72	&	99.77	\\
$|\delta d|$ (pm)	        & 1.7 	        &	0.5 	&	0.2 	&	1.1 	\\
\multicolumn{5}{c}{MBE(3)-OSV-MP2 $l_\mathrm{osv}=10^{-4},l_\mathrm{2b}=10^{-2},l_\mathrm{3b}=0.1$}\\
$\tilde{N}_{2b}~^f$	&	1772	&	2003	&	3213	&	2670	\\
$\tilde{N}_{3b}~^g$	&	9886	&	12107	&	19069	&	13852	\\
$|\delta E_c|$ (\%)	&	99.80	&	99.69	&	99.76	&	99.80	\\
$|\delta d|$ (pm)		&	1.4 	&	0.4 	&	0.2 &	0.8\\[0.5 em]
\multicolumn{5}{p{15cm}}{{\footnotesize$^a$Number of atoms. $^b$Number of
orbital basis functions. $^c$Number of auxiliary fitting functions. $^d$Number
of full 2b clusters. $^e$Number of full 3b clusters. $^f$Number of selected 2b
clusters. $^g$Number of selected 3b clusters.
}}\\
\hline
\hline
\end{tabular}
\end{table}

\subsubsection{Molecular dynamics simulation}

\begin{figure}[H]
  \begin{minipage}[t]{8cm}
  \includegraphics[width=8cm,trim={1.5cm 0 1.2cm 0},clip]{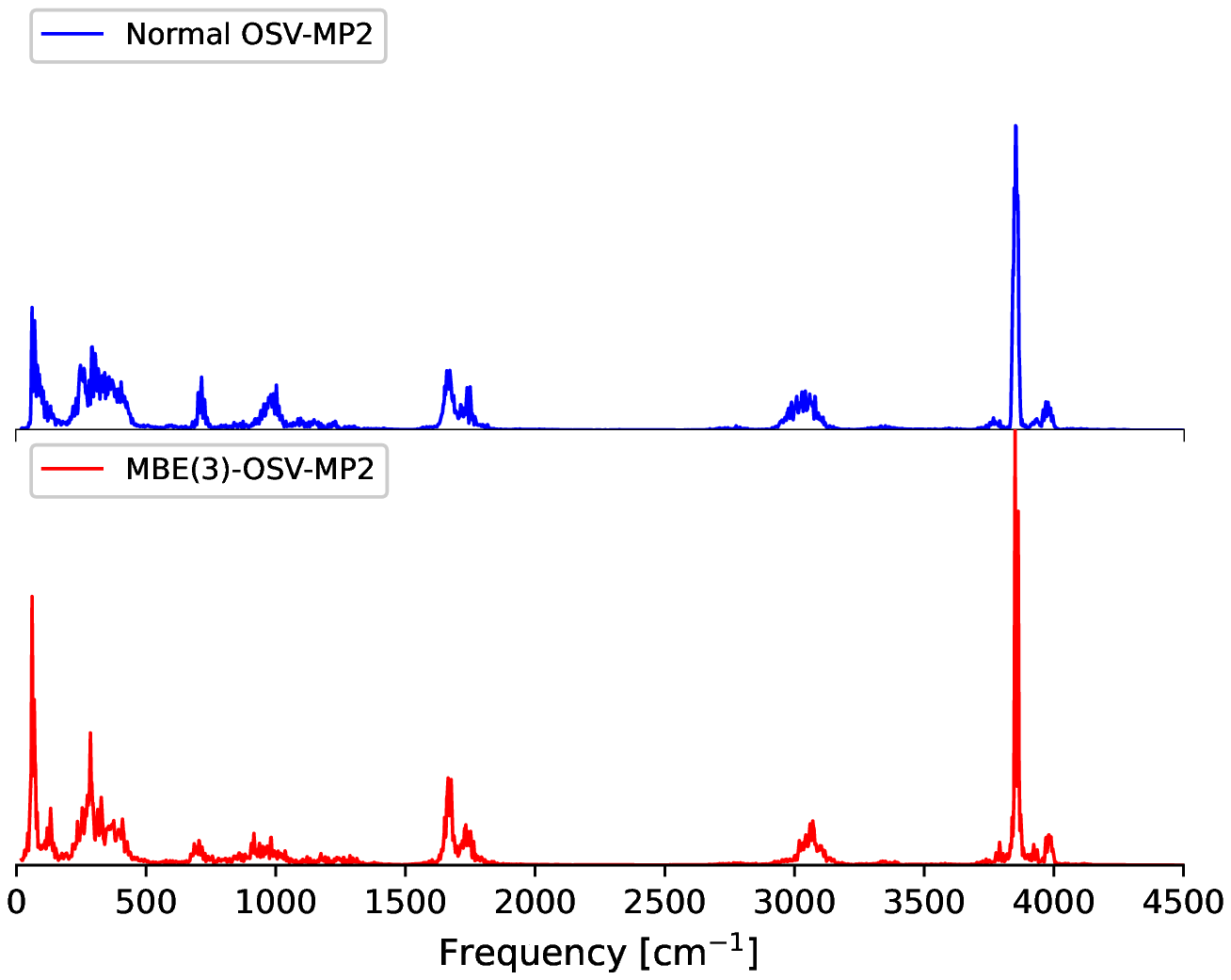}
  \caption*{(a) Eigen H$_{9}$O$_{4}^{+}$}
\end{minipage}%
\begin{minipage}[t]{8cm}
  \includegraphics[width=8cm,trim={1.5cm 0 1.2cm 0},clip]{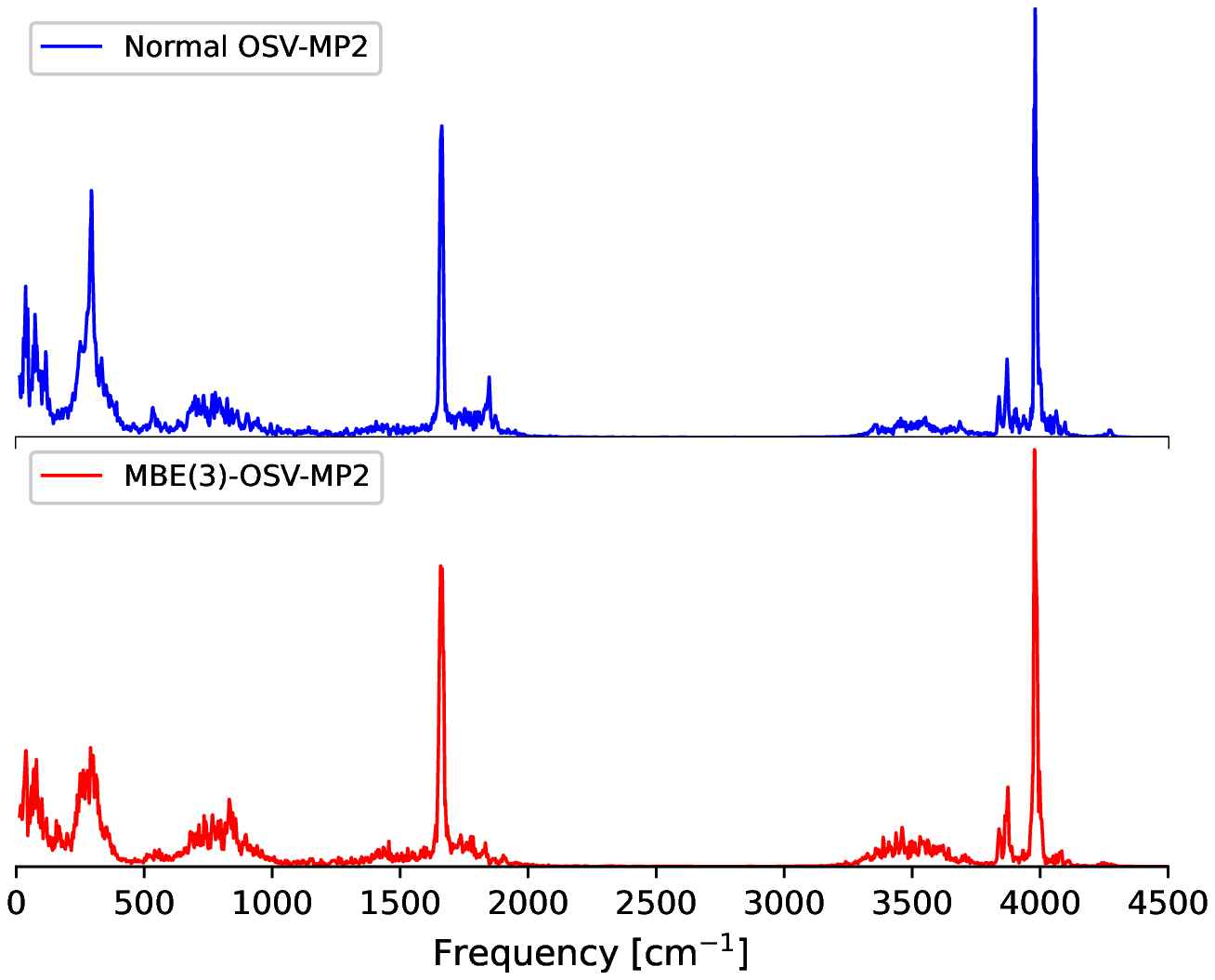}
  \caption*{(b) Zundel H$_{13}$O$_{6}^{+}$}
\end{minipage}
  \caption{Comparison of VDOS spectra between MBE(3)-OSV-MP2 and normal OSV-MP2
implementations for Eigen (a) and Zundel (b) clusters. VDOS spectra were
computed by taking the fast Fourier transform of the velocity auto-correlation
function.}
\label{fig:f_vdos_water}
\end{figure}

\begin{figure}[H]
  \centering
  \begin{minipage}[b]{8cm}
  \centering
  \includegraphics[width=6cm,trim={1.5cm 0 1.2cm 0},clip]{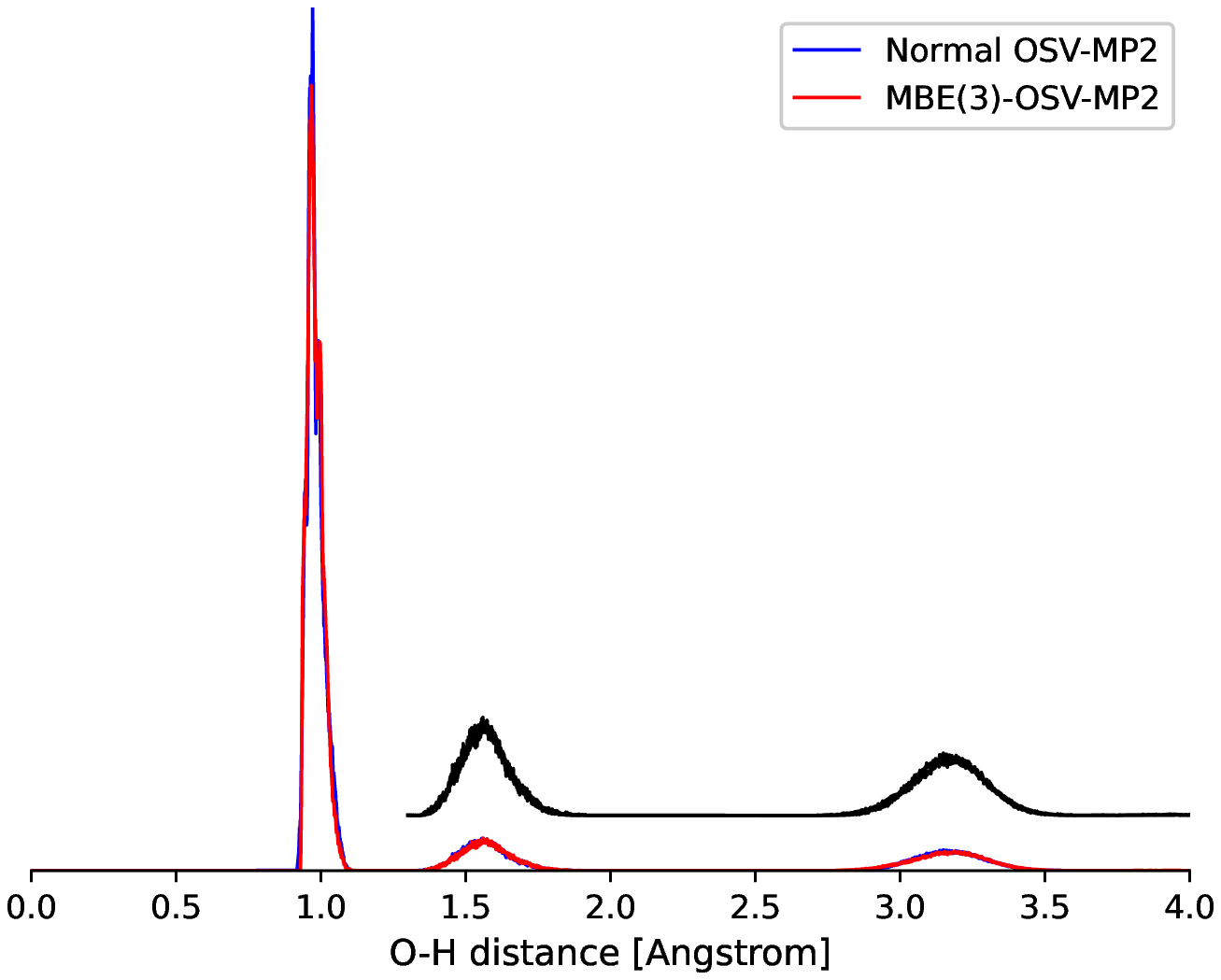}
  \caption*{(a) O-H H$_{9}$O$_{4}^{+}$}
\end{minipage}%
\begin{minipage}[b]{8cm}
  \centering
  \includegraphics[width=6cm,trim={1.5cm 0 1.2cm 0},clip]{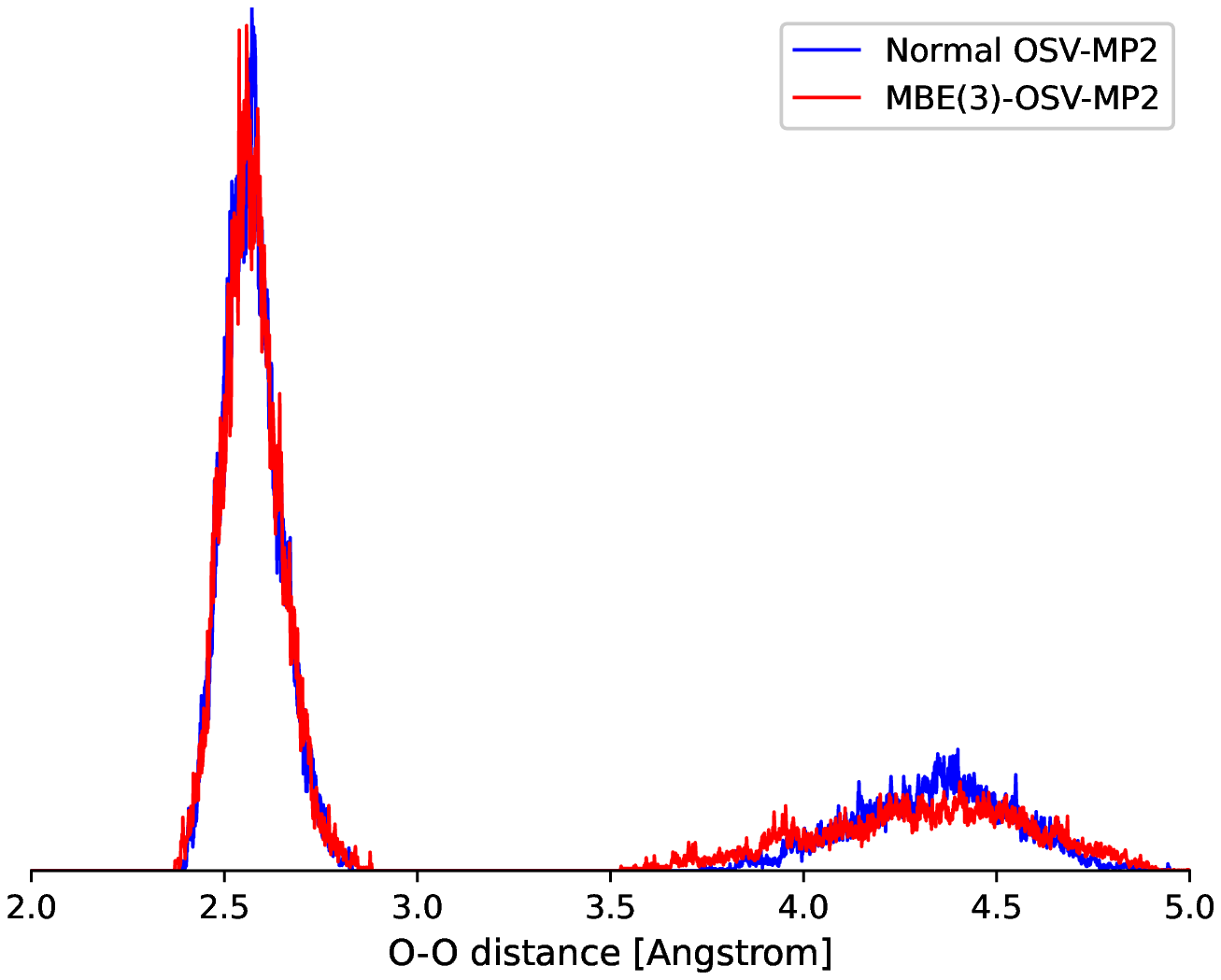}
  \caption*{(b) O-O H$_{9}$O$_{4}^{+}$}
\end{minipage}
\begin{minipage}[b]{8cm}
  \centering
  \includegraphics[width=6cm,trim={1.5cm 0 1.2cm 0},clip]{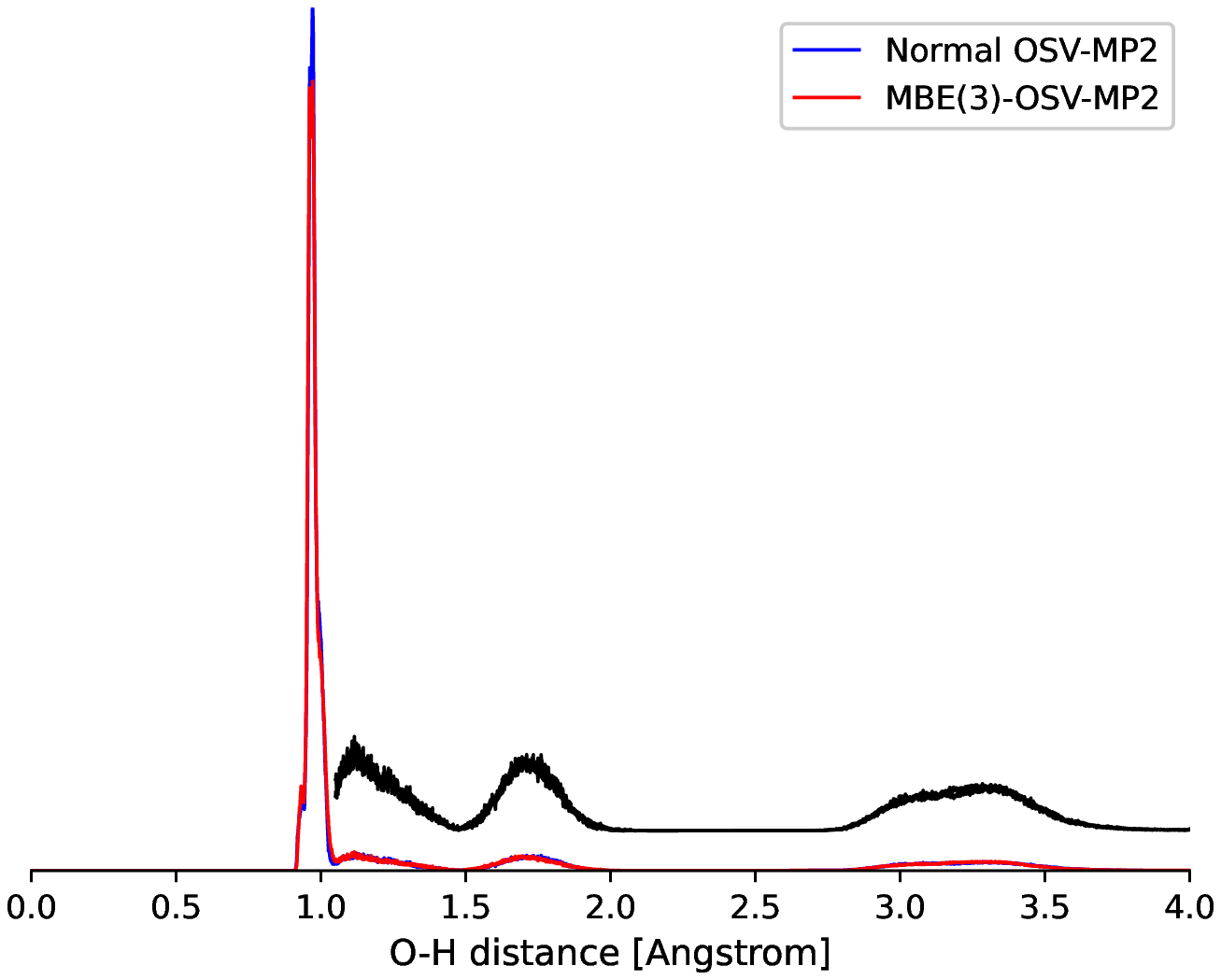}
  \caption*{(c) O-H H$_{13}$O$_{6}^{+}$}
\end{minipage}%
\begin{minipage}[b]{8cm}
  \centering
  \includegraphics[width=6cm,trim={1.5cm 0 1.2cm 0},clip]{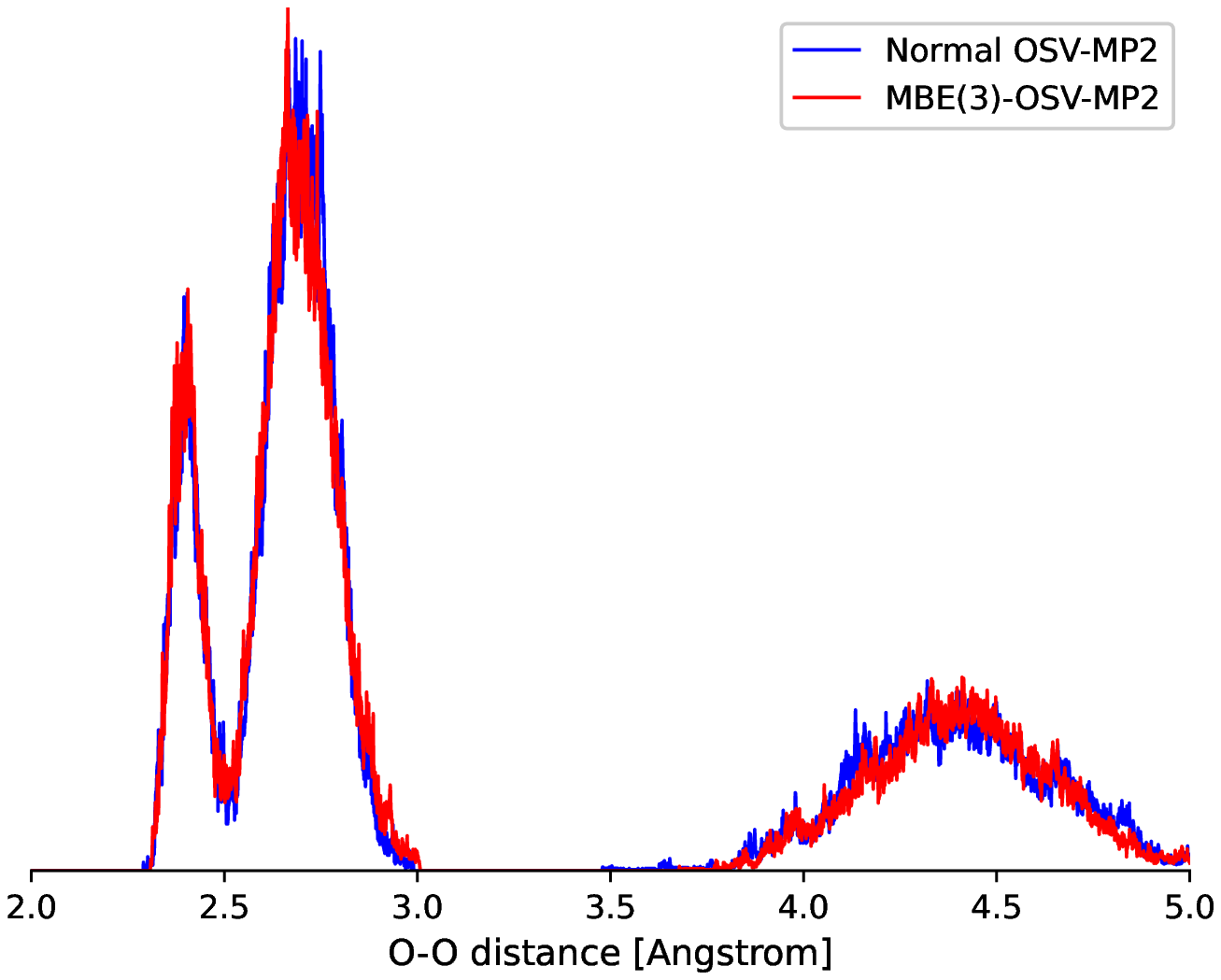}
  \caption*{(d) O-O H$_{13}$O$_{6}^{+}$}
\end{minipage}
  \caption{RDF Comparison for O-H (left) and O-O (right) distances for Eigen (a,
b) and Zundel (c, d) clusters between MBE(3)-OSV-MP2 and normal OSV-MP2
implementations. The weak  O-H spectra above 1.0 {\AA} are magnified in black,
showing O-H distances associated with hydrogen bonding and non-bonding
distribution. RDFs were prepared for full trajectories using VMD
program~\cite{humphrey1996vmd}.}
\label{fig:f_rdf_water}
\end{figure}

In our previous work~\cite{zhou2019complete}, we demonstrated that OSV-MP2
permit accurate molecular dynamics (MD) simulations that would be promising for
obtaining long-time trajectories at MP2 level of electron correlation.
For protonated water tetramer (Eigen, H$_{9}$O$_{4}^{+}$) and hexamer (Zundel,
H$_{13}$O$_{6}^{+}$) which have been often used to benchmark MD accuracy,
the OSV-MP2 method leads to accurate landscapes of the O-O/O-H radial
distribution function (RDF) and vibrational density of states (VDOS) with all
major peaks well replicated using a normal OSV selection
($l_\textrm{osv}=10^{-4}$) compared to RI-MP2 benchmark.
Here we further investigate the reliability of these MD properties derived from
MBE(3)-OSV-MP2 gradients using selected 2b and 3b clusters to propagate
classical MBE(3)-OSV-MP2/NVE trajectories for 10 ps in numerical time
integration at an interval of every 0.5 fs using the i-PI
software~\cite{kapil2019pi}. As a result, the MBE(3)-OSV-MP2/6-31+g(d,p) MD
simulation leads to energy drifts of 1.3 kJ/mol and 1.1 kJ/mol for the
protonated Eigen (H$_9$O$^{+}_{4}$) and Zundel (H$_{13}$O$^{+}_{6}$) clusters,
respectively, which are greater than the corresponding normal OSV-MP2 results
(0.0 and 0.1 kJ/mol) without using MBE(3). The energy drift, which measures the
energy conservation property of NVE simulation deviated from the linear
least-square fit to the trajectory at all time steps, reflects that the error
propagation due to the OSV and cluster selections is still within chemical
accuracy. The increased energy drift for MBE(3)-OSV-MP2 does not necessarily
alter the VDOS (Figure~\ref{fig:f_vdos_water}) and RDF
(Figure~\ref{fig:f_rdf_water}) spectra beyond statistical variance, and
all VDOS and RDF features are retrieved from MBE(3)-OSV-MP2 MD simulations, as
compared to those of normal OSV-MP2. 

\subsection{\label{sec:mbeorder}High-order MBE($n$) ($n>3$) Contribution}

MBE(3)-OSV-MP2 correlation energies and gradients demonstrated to Baker
molecules in Figure~\ref{fig:thres_sel} disclose the importance of 3b
contributions, which increases as molecular size increases. Neglect of 3b
clusters apparently results in unacceptable errors of both energy and gradient
relative to RI-MP2. Figure~\ref{fig:thres_sel} also seems to suggest
that the higher-order MBE($n$)~($n>3$) contributions beyond the 3-body level of
correlation are minor.  For dynamical properties of protonated water tetramer
and hexamer, the higher-order errors in the MBE(3)-OSV-MP2/NVE simulation are
only marginally larger than $1.0$ kJ/mol, close to chemical accuracy, and do not
make meaningful changes to the landscapes of O-O/O-H VDOS and RDF.  The
insignificance of higher-order contributions can avoid a vast number of distinct
MBE($n$) ($n>3$) clusters in otherwise catastrophic nonlinear growth with system
size that presents undesired challenges in handling efficient cutoffs of them. 

\begin{figure}[H]
  \begin{minipage}[b]{8cm}
  \includegraphics[width=8cm]{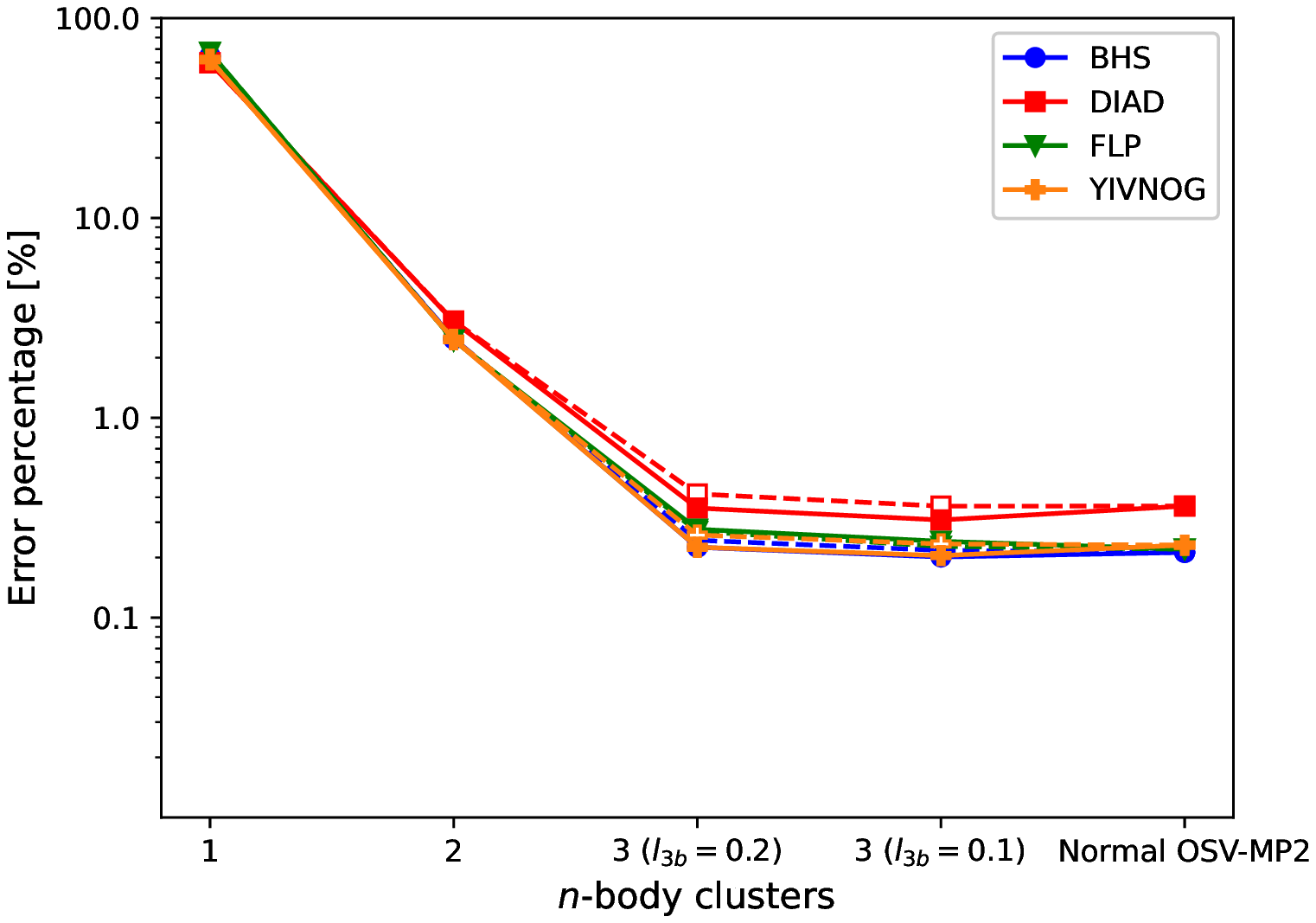}
  \caption*{(a) $l_\text{osv}=10^{-4}$, $l_\text{2b}=10^{-2}$}
\end{minipage}%
\begin{minipage}[b]{8cm}
  \includegraphics[width=8cm]{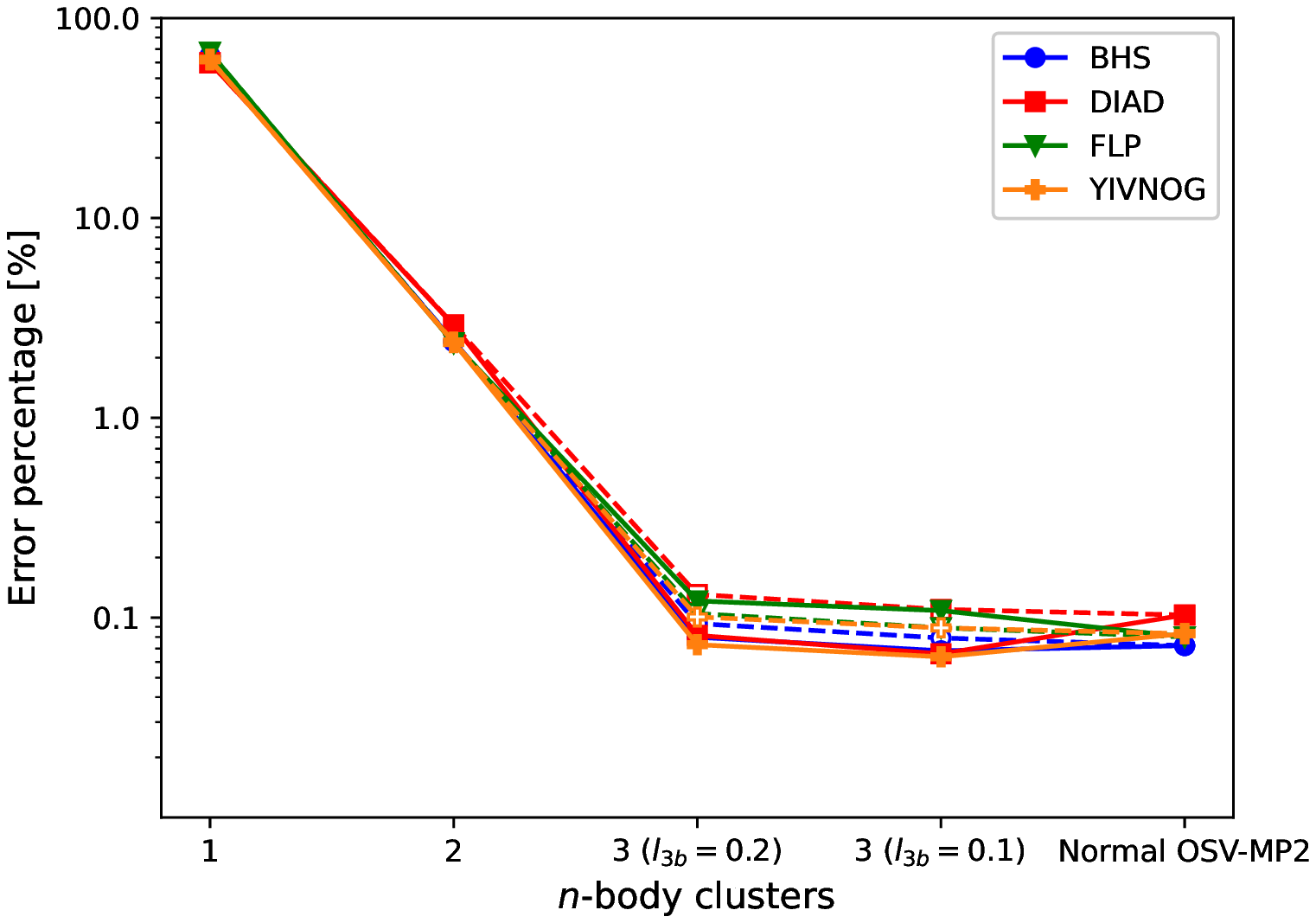}
  \caption*{(b) $l_\text{osv}=10^{-4.5}$, $l_\text{2b}=10^{-2}$}
\end{minipage}

\begin{minipage}[b]{8cm}
  \includegraphics[width=8cm]{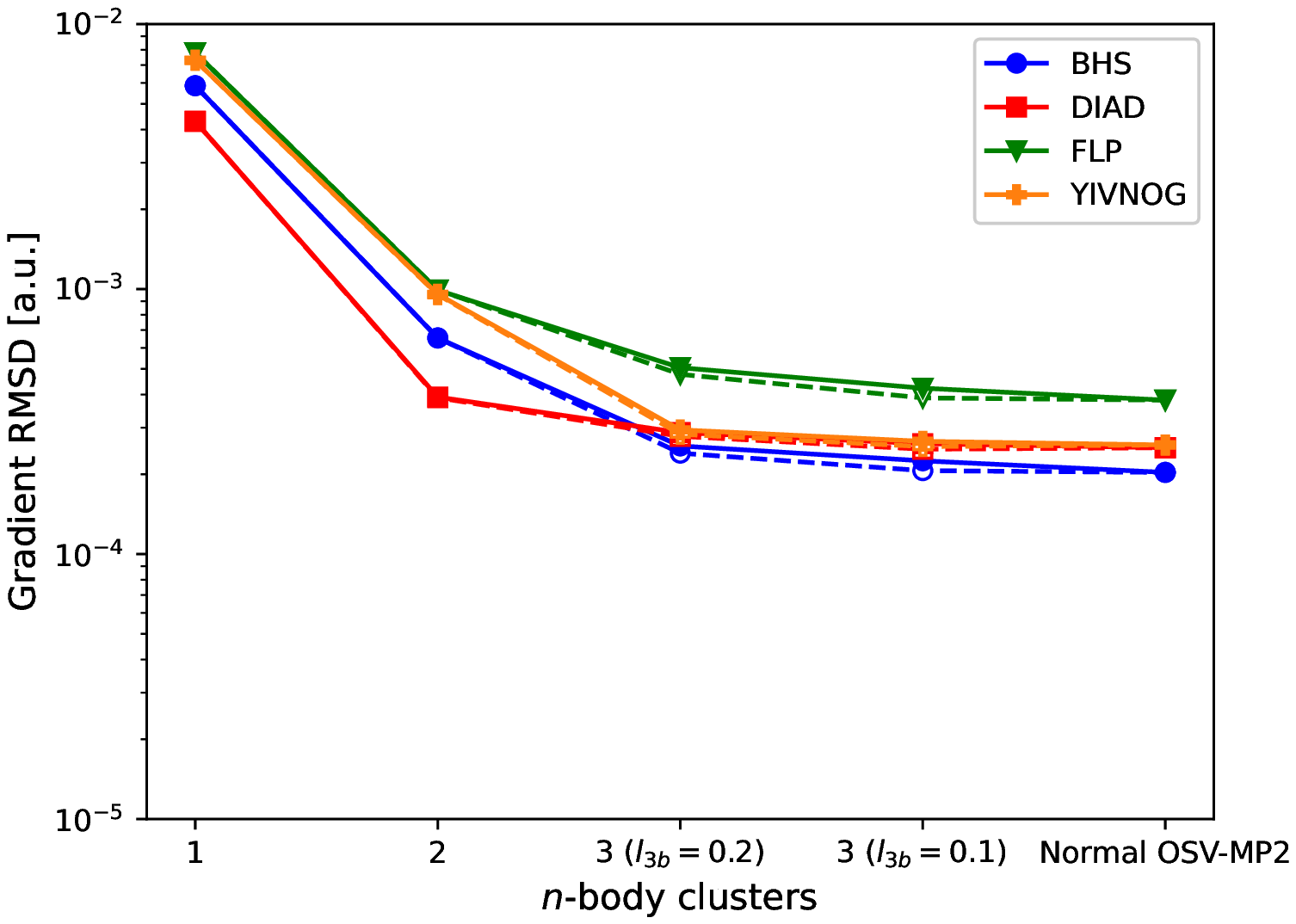}
  \caption*{(c) $l_\text{osv}=10^{-4}$, $l_\text{2b}=10^{-2}$}
\end{minipage}%
\begin{minipage}[b]{8cm}
  \includegraphics[width=8cm]{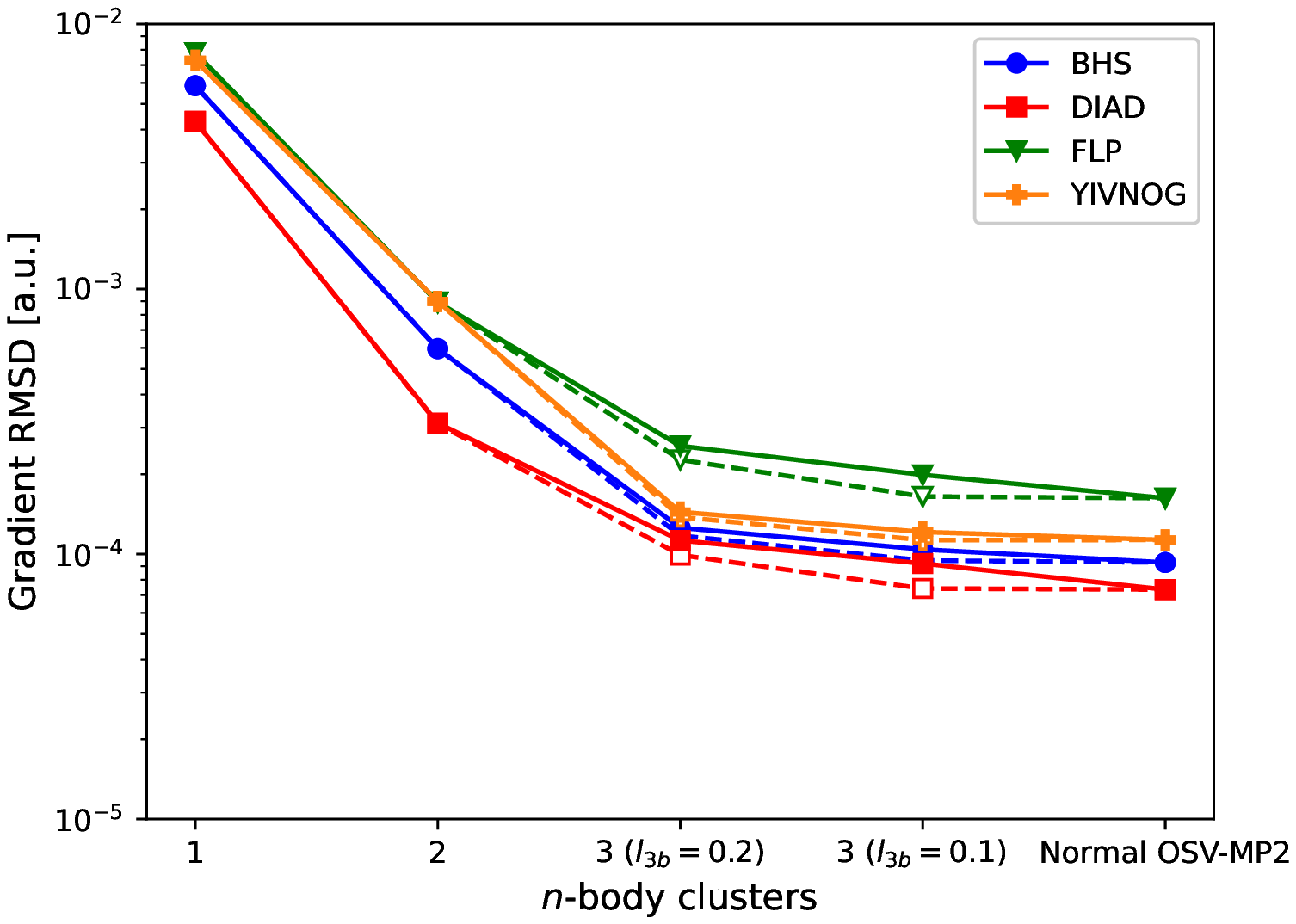}
  \caption*{(d) $l_\text{osv}=10^{-4.5}$, $l_\text{2b}=10^{-2}$}
\end{minipage}
  \caption{Errors in 1b, 2b and 3b clusters to MBE(3)-OSV-MP2 correlation energy
(a and b: percentage) and gradient (c and d: RMSD) from those of RI-MP2, for
BHS, DIAD, FLP and YIVNOG molecules computed at the def2-tzvp basis set. The
normal OSV-MP2 uses the orbital pairs from kept 2b clusters.  The high-order
one-step corrections to the collective amplitude are shown in dashed lines.}
\label{fig:mbe}
\end{figure}

To further demonstrate that the actual impact arising from MBE($n$)~($n>3$)
clusters on energy and gradient of large molecules is insignificant, we estimate
the residual error $\delta \mathbf{R}_{(ij,ij)}$ according to eq~\ref{eq:rij},
using the converged MBE(3)-OSV-MP2 collective pair amplitudes
$\mathbf{T}_{(ij,ij)}$ by which $ij$ pairs can correlate with a range of $k$ LMOs
among the union of 1b, 2b and 3b clusters, and compute the amplitude correction
$\delta \mathbf{T}_{(ij,ij)}$ in one step.  This one step posterior correction
not only couples all independent cluster amplitudes, but also correlates each
pair $ij$ with a range of close $k$ LMOs which is shown small for large molecules
using triplet-$\zeta$ basis sets, for instances, up to 13 LMOs are found
significant for BHS, DIAD, FLP, YIVNOG and C$_{60}$@catcher, and up to 10 LMOs
for (H$_2$O)$_{190}$, barely adding timing costs compared to iterative MBE(3)
residual. Figure~\ref{fig:mbe} illustrates the high-order MBE($n$) impact to
both correlation energies and gradients, using the RI-MP2 structures of BHS,
DIAD, FLP and YIVNOG molecules.  The high-order contributions appear to be
unimportant, as compared to normal OSV-MP2 results, for both OSV selections of
$l_\text{osv}=10^{-4}$ and $l_\text{osv}=10^{-4.5}$. Such corrections from the
full range of $k$ LMOs are also computed and presented in Figure~S3, which
improves both MBE(3) energies and gradients at the 3b level towards normal
OSV-MP2 results, with yet a very small magnitude within $0.05\%$ and $10^{-4}$
au, respectively.  This implies that the 3b level of cluster truncation is
indeed sufficient to achieve the accuracy of energy and gradient, close to
that of normal OSV-MP2. 

\subsection{\label{sec:mbepara}Scaling and Parallel Performance}

\begin{figure}[H]
  \centering
  \begin{minipage}{8.2cm}
  \includegraphics[width=8cm,trim={0.5cm 0 1.2cm 0},clip]{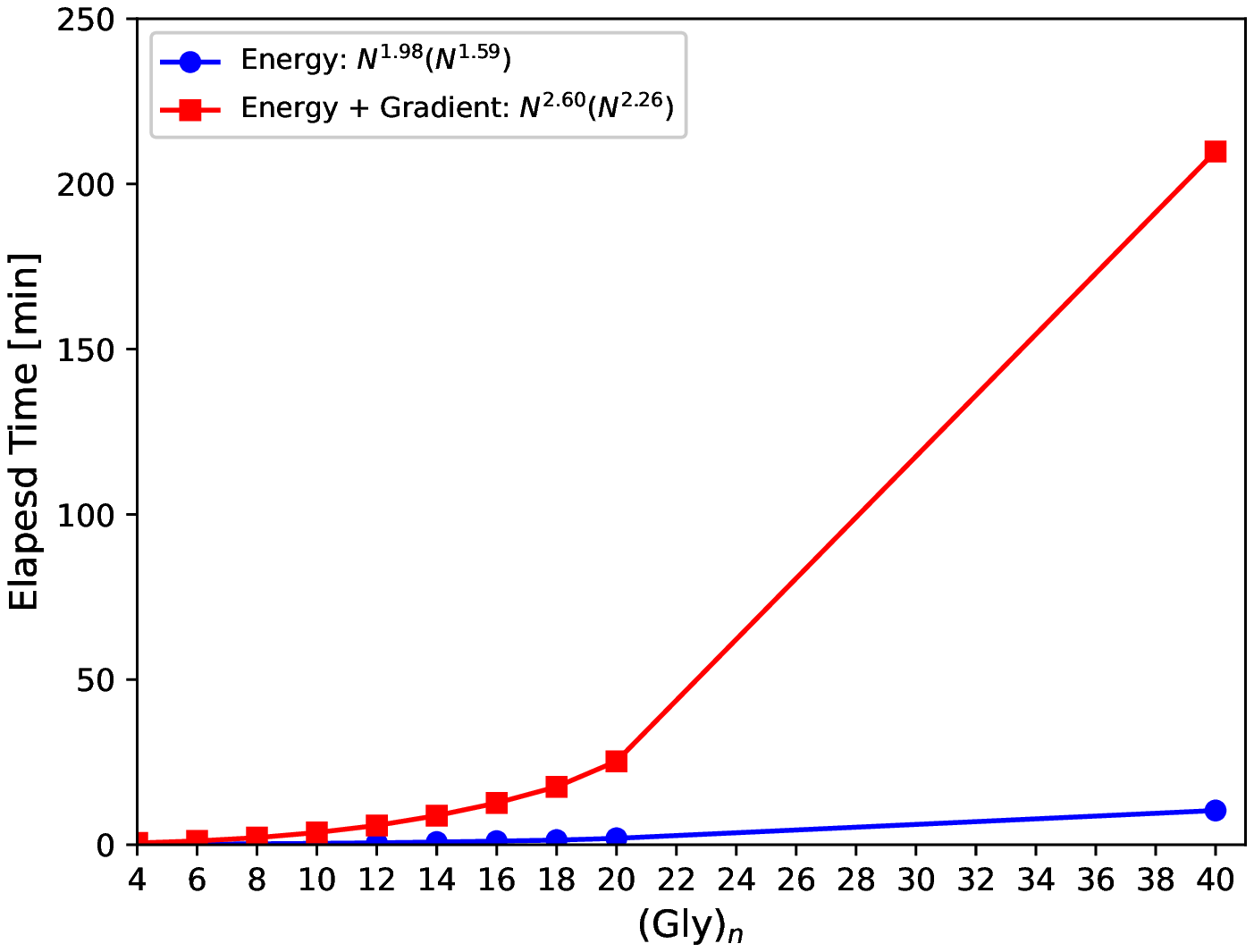}
  \caption*{(a)}
\end{minipage}%
\begin{minipage}{8.2cm}
  \includegraphics[width=8cm,trim={0.5cm 0 1.2cm 0},clip]{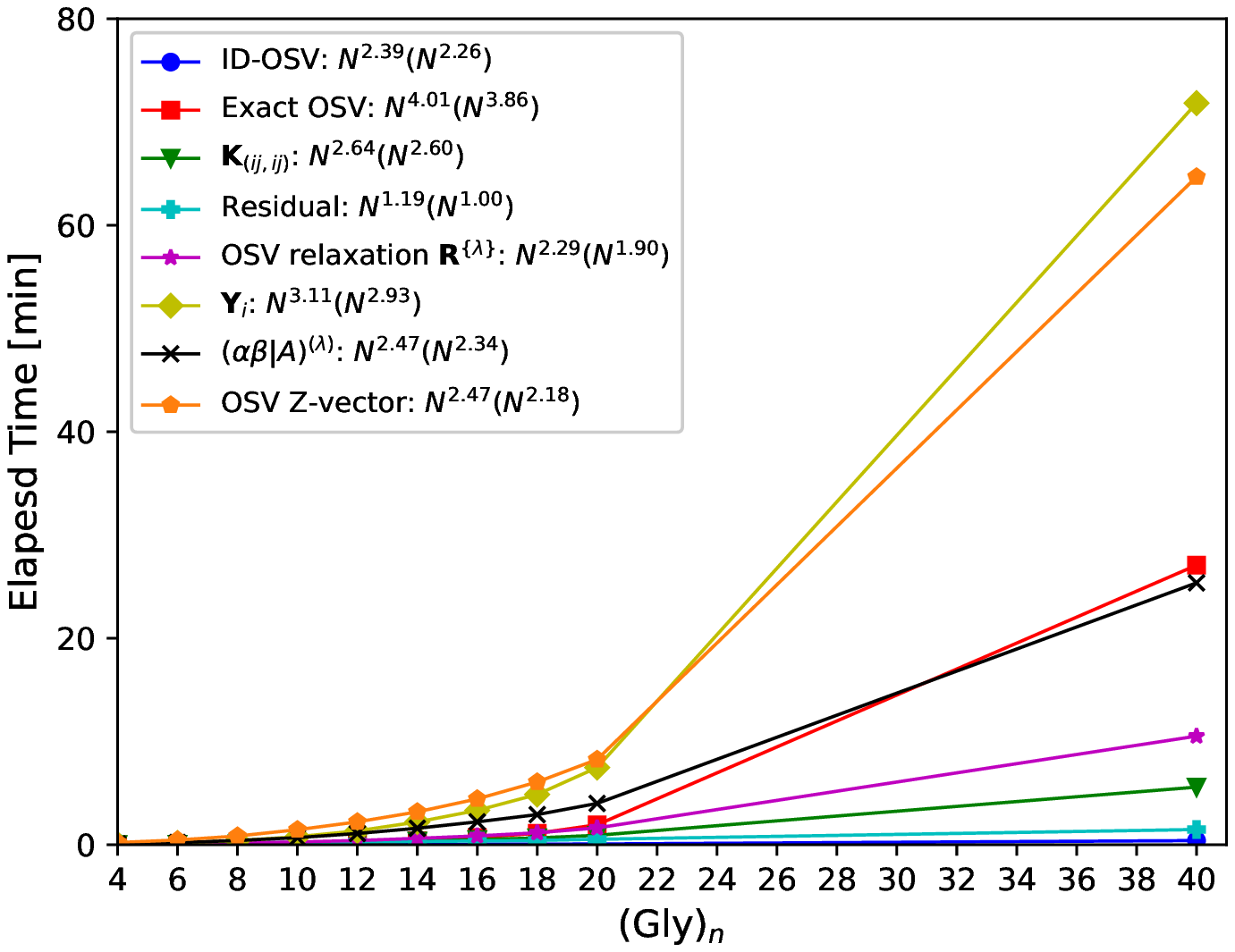}
  \caption*{(b)}
\end{minipage}
  \caption{Elapsed time of computing MBE(3)-OSV-MP2 correlation energy and
gradient (a) as well as various dominating steps (b) for polyglycine (Gly)$_{n}$
with def2-tzvp basis set on 24 CPU cores (Intel Xeon Platinum 9242@2.30GHz). The
scaling powers are presented in the legend for up to (Gly)$_{40}$ and
(Gly)$_{14}$ in the parentheses. The 3c2e half-integrals $\mathbf{J}_i$ were
computed and placed on disk. The computations of LMOs and $\mathbf{J}_i$ are not
included in MBE(3)-OSV-MP2 energy and overall scaling.} \label{fig:f_scal_size}
\end{figure}

In this section, we assess the scaling and parallel efficiency of MBE(3)-OSV-MP2
energy and gradient implementations with increasing system sizes and CPU
numbers. Polyglycine chains (Gly)$_n$ comprising up to $n=40$ units were used
for the scaling demonstration. As shown in Figure~S2, with $l_\text{2b}=10^{-2}$
and $l_\text{3b}=0.2$, the numbers of selected 2b and 3b clusters exhibit nice
linear growths with the (Gly)$_n$ lengths and are reduced by at least an order
of magnitude from the full cluster size for (Gly)$_{40}$. The overall elapsed
time of MBE(3)-OSV-MP2 energy and gradient scales according to $N^{1.59}$ and
$N^{2.26}$ up to (Gly)$_{14}$, respectively as shown in
Figure~\ref{fig:f_scal_size}a, which greatly improves the computing performance
of our previous OSV-MP2 implementation with $N^{2.74}$ for energy and $N^{2.96}$
for gradient~\cite{zhou2019complete} for similar molecular sizes. The energy and
gradient scalings increase to $N^{1.98}$ and $N^{2.60}$ towards larger
(Gly)$_{40}$, respectively, due to significantly larger half-integrals
$\mathbf{J}_i$ and $\mathbf{Y}_i$ intermediates that are stored on shared disk
and considerably increasing I/O bottleneck. Nevertheless,
although the timing cost does not scale linearly with system size, the present
implementation already allows efficient gradient computations of large molecule
containing a few hundred atoms, and meanwhile benefits fast MD simulations of
smaller molecule.  For instances, using normal cutoffs of OSVs
($l_\mathrm{osv}=10^{-4}$) and MBE(3) clusters ($l_\mathrm{2b}=10^{-2}$,
$l_\mathrm{3b}=0.2$), each single MBE(3)-OSV-MP2/def2-tzvp energy and gradient
computation of C$_{60}$@catcher complex (148 atoms) takes only 34 and 190
minutes on 24 CPUs, respectively; the MBE(3)-OSV-MP2/6-31g* MD simulation runs
on 1--2 ps trajectory length per day for porphycene molecule on 96 CPUs.

To understand the algorithmic complexities pertinent to the current
implementation, we further analyze the scaling performance of various dominating
steps within a single MBE(3)-OSV-MP2 gradient computation. As presented in
Figure~\ref{fig:f_scal_size}b, the residual time cost for amplitudes is
negligibly small and scales almost linearly with (Gly)$_n$ sizes according to
$N^{1.19}$ as a result of the linear growth of 2b and 3b clusters.  The time
complexity for OSV-specific residual relaxation $\mathbf{R}^{\{\lambda\}}$ is
shortened from $N^6$ to $N^{2.29}$ owing to massive truncations of OSVs,
discarded OSV relaxation vectors and MBE(3) clusters.  The time of generating exact OSVs
increases rapidly at $N^4$ with (Gly)$_n$ length which eventually contributes to
a large fraction of overall time expense for large molecule, but can be reduced
dramatically to $N^{2.39}$ with negligible time cost by employing approximate
ID-OSVs. The 3c2e half-transformation $\left(i\alpha\vert A\right)$ and the
evaluation of resulting OSV-based 4c2e integrals spend only moderate timings
with a scaling reduction from original $N^4$ to $N^{2.71}$ and from $N^5$ to
$N^{2.64}$, respectively, by the AO shell pair screening, sparse fitting as
well as selection of OSVs and MBE(3) clusters. The most expensive steps for
gradient computation appear to be associated with two-electron terms in, such
as, the $\mathbf{Y}_i$ intermediate ($N^{3.11}$), the Z-vector potential
($N^{2.47}$) and the derivative AO integrals $\left(\alpha\beta\vert
A\right)^{(\lambda)}$ ($N^{2.47}$), all of which add up to about half of the
overall gradient time.  This indicates that the performance for very large
molecules begins to be certainly bounded to these  predominant costs.
For energy alone, the computation of OSV two-electron integrals
$\mathbf{K}_{(ij,ij)}$ for strong 2b clusters dominates the timing cost, while
the timing of $\mathbf{K}_{(i,j)}$ for weak 2b clusters is negligible.

\begin{figure}[H]
  \begin{minipage}[b]{8cm}
   \includegraphics[width=8cm,trim={0.8cm 0 1.2cm 0},clip]{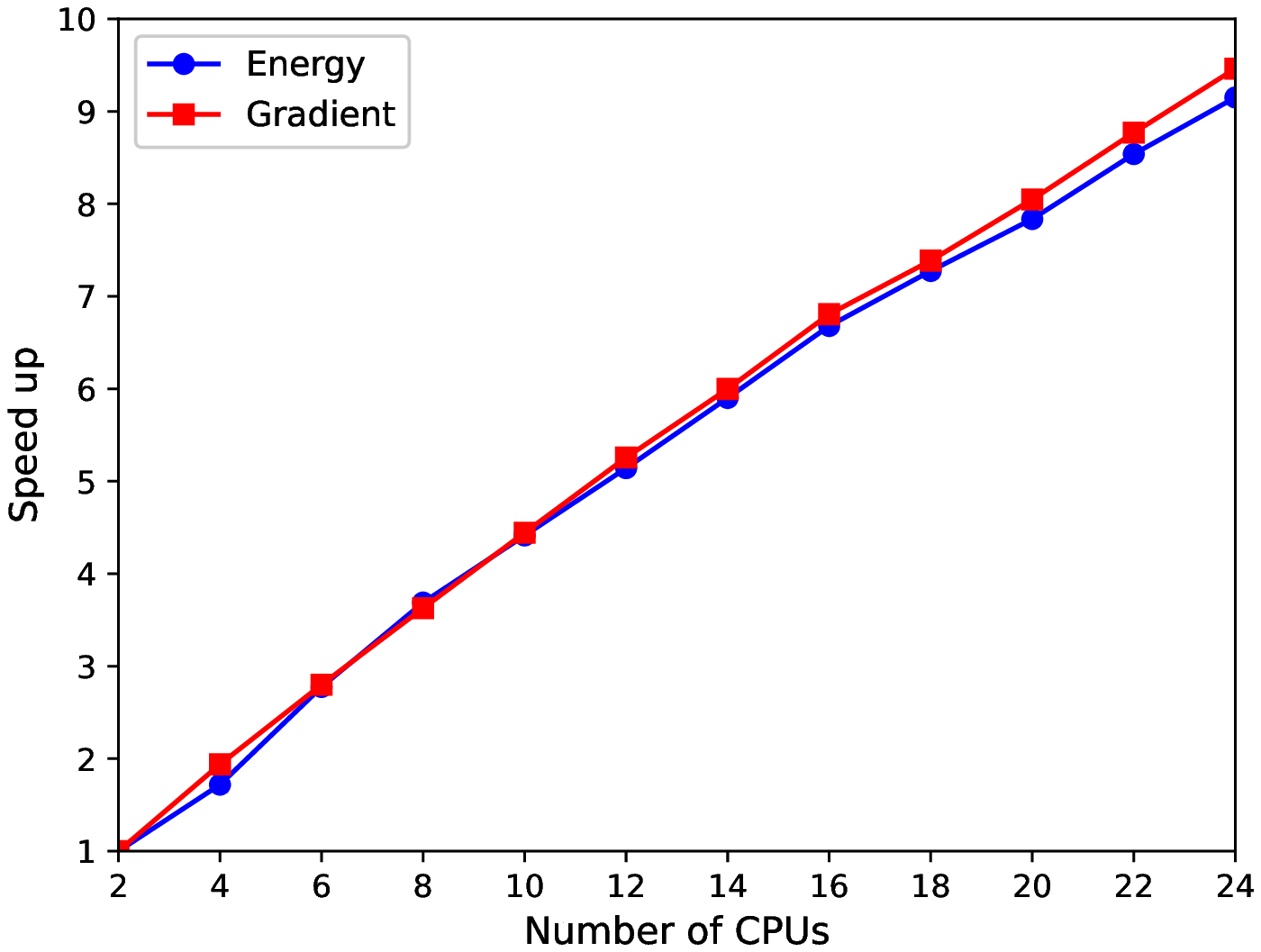}
  \caption*{(a) (Gly)$_{20}$}
  \end{minipage}
  \begin{minipage}[b]{8cm}
   \includegraphics[width=8cm,trim={0.8cm 0 1.2cm 0},clip]{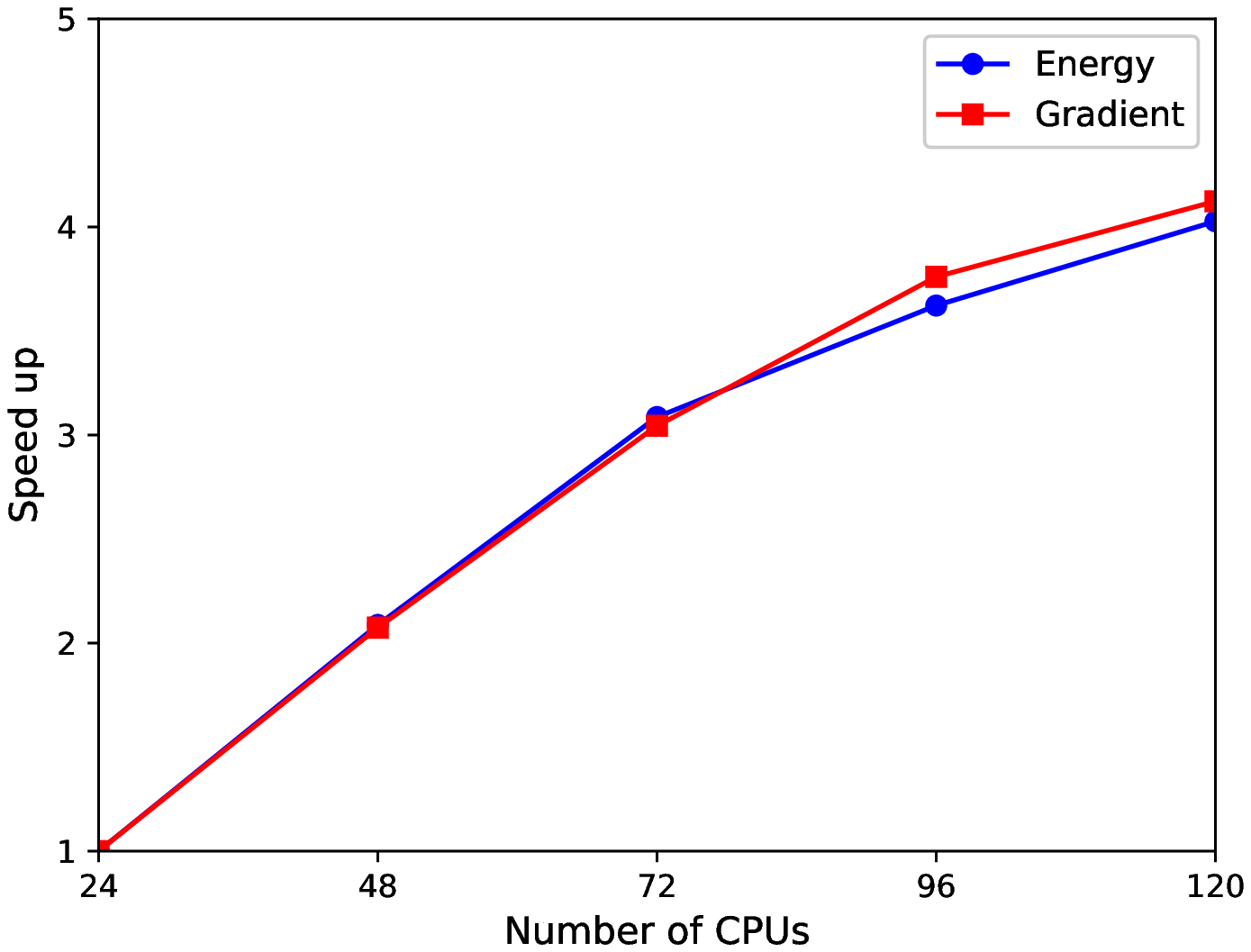}
  \caption*{(b) C$_{60}$@catcher}
  \end{minipage}%
  \caption{Parallel speedups in elapsed time of MBE(3)-OSV-MP2 energy and
gradient computations with an increasing number of CPUs (Intel Xeon Platinum
9242@2.30GHz) for (Gly)$_{20}$ (a) and 120 CPUs across 10 nodes (Intel Xeon
E5-2692 v2@2.20 GHz) for C$_{60}$@catcher (b).}
\label{fig:f_scal_core}
\end{figure}

Next, we demonstrate the parallel speedups with respect to the number of CPUs
presented in Figure~\ref{fig:f_scal_core} for (Gly)$_{20}$ and C$_{60}$@catcher
complexes. For (Gly)$_{20}$ computed on 2--24 CPUs, a satisfactory parallel
speedup of MBE(3)-OSV-MP2 gradient computation is achieved, relative to that
using 2 CPUs.  For larger C$_{60}$@catcher molecule, three- and
four-fold speedups in elapsed time are observed on 72 and 120 CPUs compared to
the timing on 24 CPUs, respectively.  Overall, a parallel scalability is nearly
100\% for a smaller number of CPUs and drops to 80\% when a large number of CPUs
is employed. Although our implementation is based on passive one-sided
communication of supposedly low synchronization latency, it is inevitable that
the number of parallel I/O disk accesses grows with increasing number of CPUs,
and more importantly, the uneven distribution of parallel tasks becomes an issue
that further adds synchronization overheads and reduces the parallel
scalability.

\begin{figure}[H]
\begin{minipage}[b]{6cm}
  \centering
  \includegraphics[width=5cm]{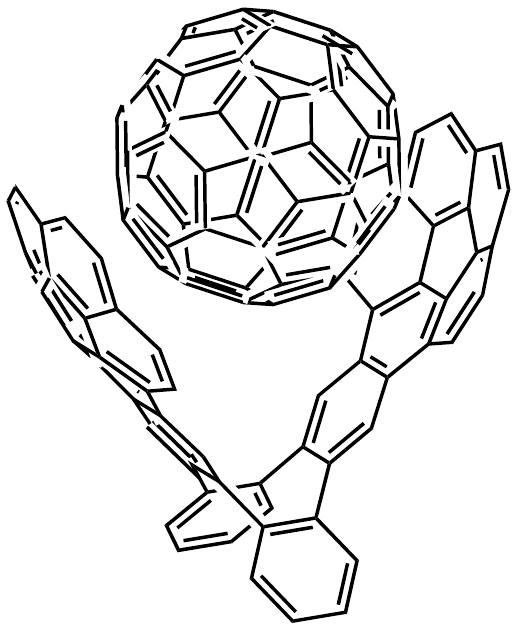}
   \caption*{C$_{60}$@catcher}
\end{minipage}
\begin{minipage}[b]{6cm}
  \includegraphics[width=5cm]{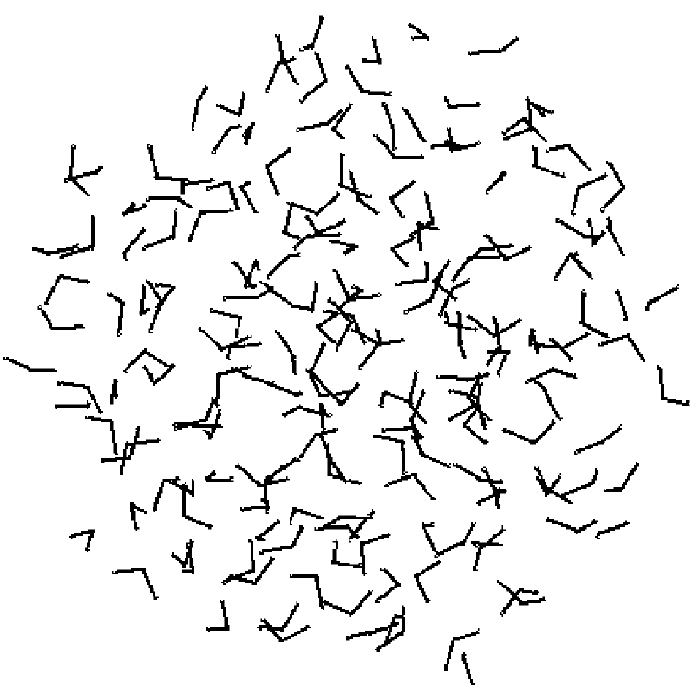}
  \caption*{(H$_2$O)$_{190}$}
\end{minipage}
  \caption{Chemical formula of large molecules for MBE(3)-OSV-MP2 application.
The coordinate of C$_{60}$@catcher is from ref.\cite{sure2015comprehensive} and
that of (H$_2$O)$_{190}$ is from ref.\cite{watergeo}.}
  \label{fig:f_mol_large}
\end{figure}

Finally, we take C$_{60}$@catcher and (H$_2$O)$_{190}$
(Figure~\ref{fig:f_mol_large}) as examples to demonstrate the performance of
parallel MBE(3)-OSV-MP2 gradient computation for large molecules. The timing
results for a single gradient computation are given in
Table~\ref{tab:t_time_large}.  The total elapsed time of the parallel energy and
gradient computations is about 190 minutes for C$_{60}$@catcher/def2-tzvp, 543
minutes for (H$_2$O)$_{190}$/vdz and 3588 minutes for (H$_2$O)$_{190}$/vtz on 24
CPUs.  The correlation energy computation of MBE(3)-OSV-MP2 alone takes about
only 30 minutes for C$_{60}$@catcher/def2-tzvp,  72 minutes for
(H$_2$O)$_{190}$/vdz and 806 minutes for (H$_2$O)$_{190}$/vtz.  Further timing
speedup can be achieved when more CPU resources become available, for instance,
there is a four-fold speedup on 120 CPUs for C$_{60}$@catcher, which makes it
now feasible to afford structure optimization for large molecules with a few
thousand orbital functions and ten thousand fitting functions in a reasonable
time.  Again, we find that the bottleneck steps still point to the computation
of $\mathbf{Y}_{i}$, the derivative integrals $(\alpha\beta \vert
A)^{(\lambda)}$ and the OSV Z-vector solution, which take the time fractions of
23.7\%, 16.2\% and 38.0\% for C$_{60}$@catcher, 18.8\%, 27.2\% and 34.4\% for
(H$_2$O)$_{190}$/vdz, as well as 21.8\%, 21.9\% and 22.2\% for
(H$_2$O)$_{190}$/vtz. Solving OSV Z-vector equation takes up the largest portion
of the MBE(3)-OSV-MP2 gradient time for both systems, as the evaluation of
$\sum_{kl}{\Lambda}_{kl}\mathcal{B}_{kl,ai}$ belonging to the Z-vector source
term $W_{ia}$ of eq~\ref{eq:sourcew} is rather inefficient in our current
implementation. It is noticed that the exact OSV generation costs 403 minutes
that is about half of the energy computational time for (H$_2$O)$_{190}$ using
triplet-$\zeta$ basis set, but this is dramatically reduced to only 22 minutes
when approximate ID-OSV (section~\ref{sec:spar}) is generated.

\begin{table}[H]
  \caption{Comparisons of elapsed time (minutes) and timing fraction (\%) in various
MBE(3)-OSV-MP2 energy and gradient steps for single-point computations of
C$_{60}$@catcher and (H$_{2}$O)$_{190}$ on 24 CPUs (Intel Xeon Platinum
9242@2.30GHz).}
  \label{tab:t_time_large}
  \begin{tabular}{ccccccc}
  \hline
\hline																	
Molecular sizes  & \multicolumn{2}{c}{C$_{60}$@catcher}	& \multicolumn{4}{c}{(H$_{2}$O)$_{190}$}\\
\hline
basis set 	 & \multicolumn{2}{c}{def2-tzvp}        & \multicolumn{2}{c}{cc-pvdz}   & \multicolumn{2}{c}{cc-pvtz} \\
atoms 		 & \multicolumn{2}{c}{148}              & \multicolumn{2}{c}{570}	& \multicolumn{2}{c}{570}     \\
orbital basis	 & \multicolumn{2}{c}{3888}             & \multicolumn{2}{c}{4560}	& \multicolumn{2}{c}{11020}   \\
MP2 fitting basis& \multicolumn{2}{c}{9540}             & \multicolumn{2}{c}{15960}	& \multicolumn{2}{c}{26790}   \\
\hline
Main steps         &	Time                            &	Fraction & Time &	Fraction &	Time &	Fraction \\
	\hline	
$(i\alpha \vert A)$ 		& 9.4 & 5.0 & 39.1 & 7.2 & 264.8 & 7.4\\
exact OSV                       & 6.6       & 3.4       & 17.6       & 3.2       & 402.9       & 11.2      \\
          (ID-OSV)              &     (1.9) &           &      (1.9) &           &       (22.5) &     \\
OSV $\mathbf{S}/\mathbf{F}$ 	& 0.3 & 0.2 & 0.3 & 0.1 & 2.8 & 0.1 \\
OSV $\mathbf{K}$ matrix 	& 6.2 & 3.3 & 8.1 & 1.5 & 128.7 & 3.6 \\
residual iteration 		& 6.0 & 3.2 & 6.1 & 1.1 & 6.4 & 0.2 \\
residual relaxation $\mathbf{R}^{\{\lambda\}}$ & 13.5 & 7.1 & 34.8 & 6.4 & 418.0 & 11.7 \\
$\mathbf{Y}_{i}$ evaluation 	& 45.1 & 23.7 & 102.4 & 18.8 & 780.6 & 21.8 \\
$(\alpha\beta|A)^{(\lambda)}$ 	& 30.7 & 16.2 & 147.8 & 27.2 & 785.8 & 21.9 \\
OSV Z-vector 			& 72.2 & 38.0 & 187.0 & 34.4 & 797.6 & 22.2 \\
\hline         
total$~^a$			& 189.9 & 100  & 543.3 & 100 &  3587.7 & 100\\ [0.5em]
\multicolumn{7}{p{15cm}}{\footnotesize$^a$The total elapsed time is based on
the exact OSV generation.}\\
\hline																	
\hline																	
  \end{tabular}
\end{table}

\section{ILLUSTRATIVE APPLICATIONS}

\begin{figure}[H]
  \includegraphics[width=10cm]{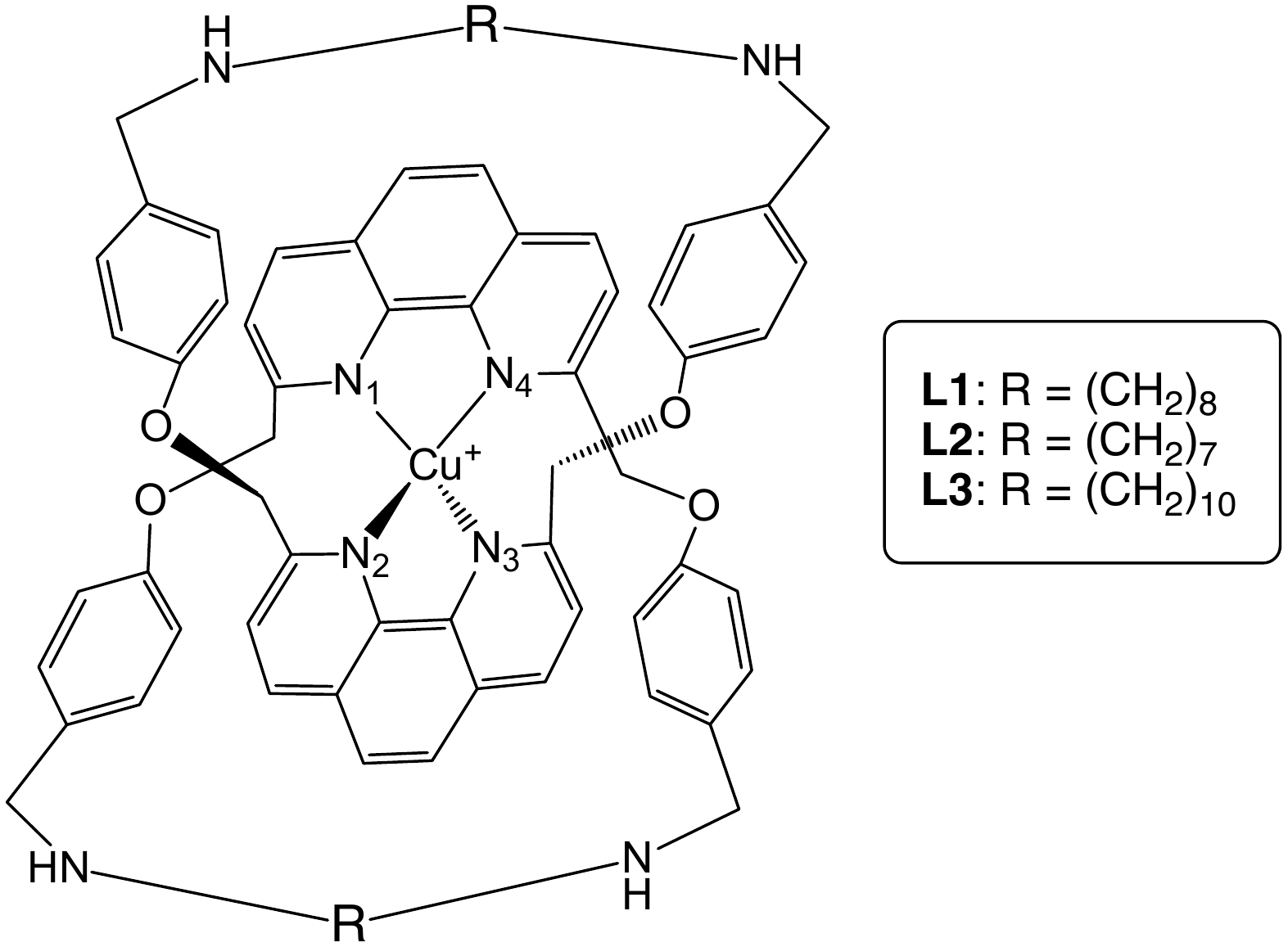}
  \caption{Chemical formula of Cu(I)-catenane complex interlocking L1, L2 and
L3 ligands, respectively.}
  \label{fig:f_mol_cu}
\end{figure}

We showcase two brief applications of MBE(3)-OSV-MP2 gradient implementation to
illustrate: (i) the variation of mechanical bond length for tuning catalytic
activity of Cu(I) complex supported by interlocked catenane
ligands\cite{zhu2020cross} (Figure~\ref{fig:f_mol_cu}), and (ii) the N-H
vibrational signature associated with $cis$/$trans$ tautomerization due to
double hydrogen transfer in porphycene molecule~\cite{litman2019elucidating}
from the MP2-level electron correlation and classical protons. Both systems
demand tremendous tasks in computing analytical energy gradients for structure
optimization and MD evolution, which are rather expensive using conventional MP2
method.

\subsection{\label{sec:catenane}Cu(I)-Catenane Interlocking Coordination Structure}
The tetradentate Cu(I) complex mechanically interlocking catenane ligands has
been recently demonstrated to selectively catalyze C(sp\textsuperscript{3})--O
dehydrogenation between phenol and bromodicarbonyl\cite{zhu2020cross}.  It was
found that different catenane topologies and peripheral lengths effectively
managed Cu(I)-catenane mechanical bonds by adjusting Cu(I) coordination
environment, leading to varying catalytic activity for a broad scope of
substrates.  The dehydrogenative coupling of phenol and diethyl bromomalonate
reveals experimentally that the Cu(L1) and Cu(L3) complexes in relatively loose
mechanical bonds with long L1 (R=(CH$_2$)$_{8}$) and L3 (R=(CH$_2$)$_{10}$)
ligands have a high catalytic activity in nearly 77--80\% product yield, while
Cu(L2) complex in tight bond with short L2 (R=(CH$_2$)$_{7}$) considerably
compromises the product generation at a yield of only 52\%. 

\begin{table}[H]
  \caption{The distances of N-Cu bonds (pm) and the dihedral angles
(\textdegree) between the N$_1$-Cu-N$_2$ and N$_3$-Cu-N$_4$ planes for
[Cu(L1)]PF$_{6}$, [Cu(L2)]PF$_{6}$ and [Cu(L3)]PF$_{6}$ optimized by
MBE(3)-OSV-MP2/def2-tzvp (all electrons) and B3LYP-D3BJ/Lanl2dz/6-31g(d,p). The
DFT values were obtained using ORCA software package\cite{neese2018software}.
$V_\text{coor}$ is the volume (pm$^3$) of the coordination cavity which is
measured as the Cu-centred polyhedral volume enclosed with four nitrogen
vertices.}
  \label{tab:t_cu}
  \begin{tabular}{llrrr}
\hline\hline										
	Method	&&	 [Cu(L1)]PF$_{6}$	&	[Cu(L2)]PF$_{6}$	&	[Cu(L3)]PF$_{6}$	\\
\hline		 							
MBE(3)-OSV-MP2	&N$_1$-Cu       &	202.22	&	200.22	&	201.75	\\
		&N$_2$-Cu       &	197.53	&	197.44	&	198.26	\\
		&N$_3$-Cu       &	202.30	&	200.51	&	201.39	\\
		&N$_4$-Cu       &	197.47	&	197.12	&	198.23	\\
		&dihedral angle &	103.96	&	104.76	&	105.23	\\
		&$V_{\text{coor}}$ 	&	3377975.78	&	3346818.43	&	3366231.91	\\
		&$\Delta V_{\text{coor}}$&	0.00	&	-31157.35	&	-11743.87	\\
\hline									
B3LYP-D3BJ	&N$_1$-Cu&205.48	&	204.50	&	205.52	\\
		&N$_2$-Cu&204.99	&	208.76	&	208.21	\\
		&N$_3$-Cu&205.48	&	204.01	&	205.58	\\
		&N$_4$-Cu&205.00	&	209.68	&	207.53	\\
		&dihedral angle &	107.77	&	105.80	&	107.86	\\
		&$V_{\text{coor}}$ &	3595533.80	&	3655158.99	&	3626158.36	\\
		&$\Delta V_{\text{coor}}$ &	0.00	&	59625.19	&	30624.56	\\
\hline\hline													
  \end{tabular}
\end{table}

The structures of [Cu(L1)]PF$_{6}$,  [Cu(L2)]PF$_{6}$ and [Cu(L3)]PF$_{6}$ were
optimized at the MBE(3)-OSV-MP2/def2-tzvp level of theory with all electrons
correlated. As shown in Table \ref{tab:t_cu}, the small change in the number of
methylene groups does not make a large impact on the Cu-N distances, nor on the
dihedral angles, causing less than 2 pm deviations in Cu-N lengths and 2 degrees
in dihedral angles, which can be however  consistently distinguished by the
MBE(3)-OSV-MP2 method.  As shown in Table \ref{tab:t_cu}, the polyhedral volume
of [Cu(L1)]PF$_{6}$ with medium size ligand is larger than those of both
[Cu(L3)]PF$_{6}$ with the longest ligand by 0.5\% and [Cu(L2)]PF$_{6}$ with the
shortest ligand by 0.9\%. The $V_\text{coor}$ ordering $\text{[Cu(L1)]P}_6 >
\text{[Cu(L3)]P}_6 > \text{[Cu(L2)]P}_6$ for three catenane ligands agrees with
the ranking of their catalytic efficacy by experiments.  This indicates that the
coordination space accommodating the Cu ion can be adjusted by tuning the ligand
length, which creates an open and responsive coordination environment for
substrates.  However, the Cu coordination volume does not scale proportionally
with the length of the catenane ligand due to intricacies of Cu-catenane
interlocking interaction and ligand topology. In passing, the structures from
MBE(3)-OSV-MP2 are also compared with DFT/B3LYP-D3BJ results. While the
MBE(3)-OSV-MP2 predicts a reduction of Cu coordination volume from
$\text{[Cu(L1)]P}_6$, the B3LYP-D3BJ results are opposite and give a volume
expansion with a different ordering $\text{[Cu(L2)]P}_6 > \text{[Cu(L3)]P}_6 >
\text{[Cu(L1)]P}_6$ from that of MBE(3)-OSV-MP2. Our results suggest that these
subtle structure changes are indeed susceptible to different correlated energy
and gradient models which are critically essential.

\subsection{\label{sec:porphycene}Porphycene Tautomerization from MD Simulation}

\begin{figure}[H]
\subfloat[\label{fig:f_rdf_por}]{
  \includegraphics[width=0.5\textwidth]{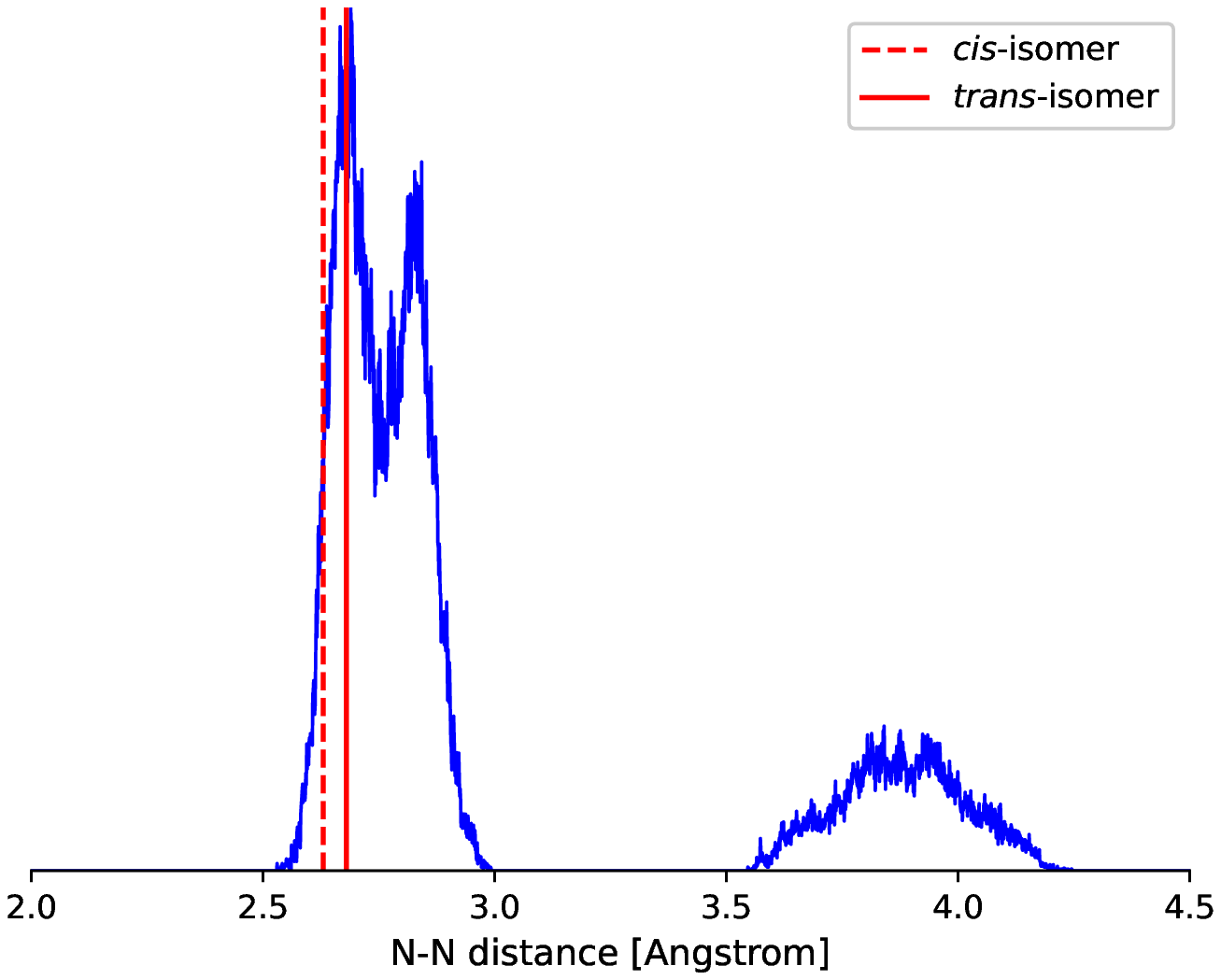}
}
\subfloat[\label{fig:f_vdos_por}]{
  \includegraphics[width=0.5\textwidth]{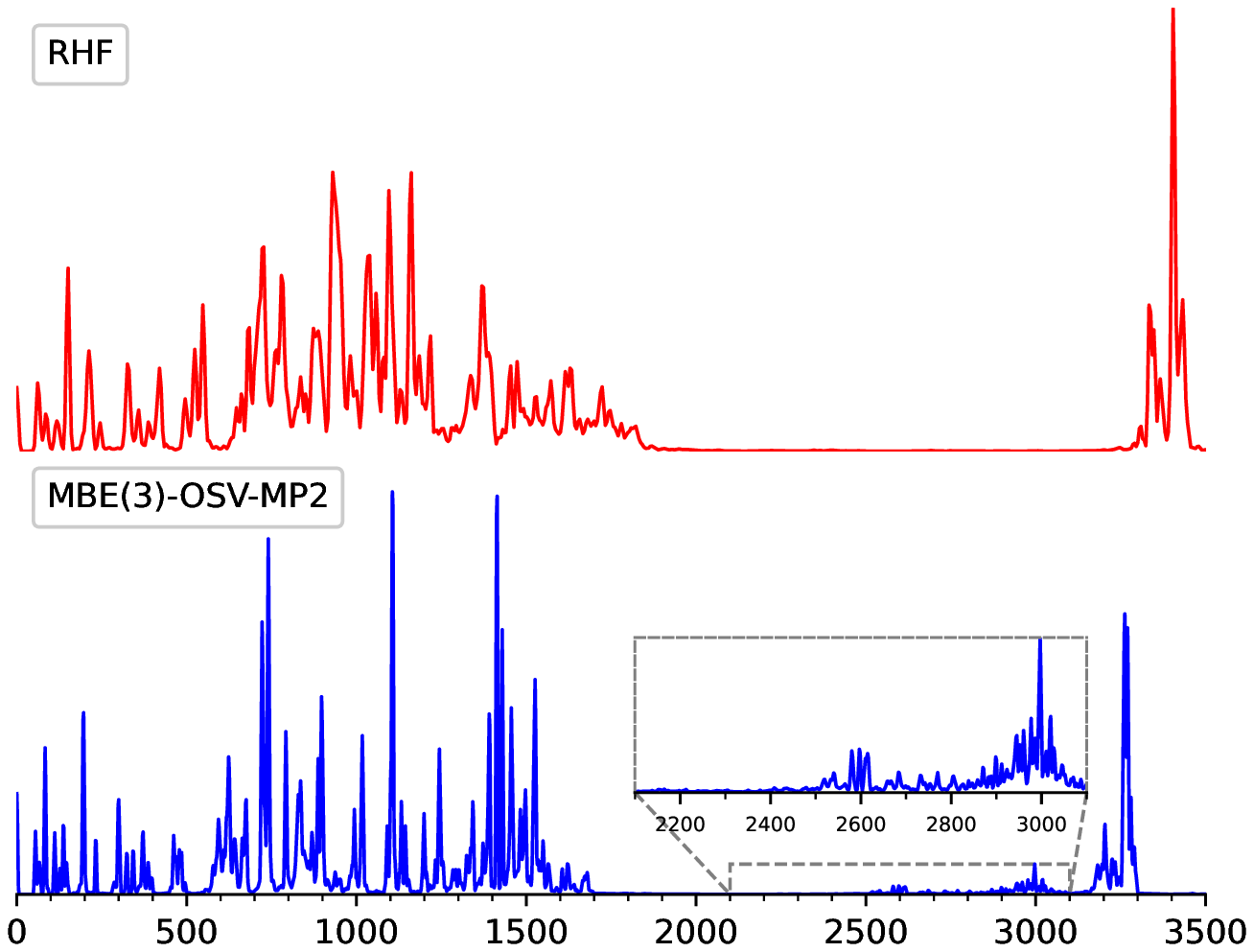}
}
  \caption{The MBE(3)-OSV-MP2/6-31g*/MD simulation from a single  trajectory for
porphycene: the N-N RDF (a) and VDOS (b). The red vertical lines label the
distances of equilibrium $trans$-Pc (solid) and  $cis$-Pc (dashed). The 10 ps
trajectory was generated at a time step of 0.5 fs from the thermostated NVT
simulation at the average temperature of 291.9 K (the temperature drift was 1.5
K), followed by another 10 ps NVE run leading to the computed RDF and VDOS
results.  The MBE(3)-OSV-MP2/MD simulation was carried out on 96 CPUs (Intel
Xeon Platinum 9242@2.30GHz).}
\end{figure}

Porphycene (Pc, C$_{20}$H$_{14}$N$_4$) is a complex prototypical molecule in
which a fast double hydrogen transfer (HT) is believed to occur at room
temperature along the strong intramolecular hydrogen bonds in the molecular
cavity formed by four
nitrogens\cite{gawinkowski2012vibrations,litman2019elucidating}. This HT-based
tautomerization reaction proceeds via an internal
$\mathrm{N}-\mathrm{H}\cdots\mathrm{N}$ pathway resulting in different
tautomeric isomers: \textit{cis}-Pc tautomer where two hydrogens are bonded to
nitrogens on the same side and \textit{trans}-Pc tautomer with two hydrogens
connected to nitrogens on the other side. However, the standard harmonic
frequency calculation assigns only a strong single peak around 2900 cm$^{-1}$ to the N-H
stretching vibration, while experimental infrared (IR) spectrum shows a
significantly broadened and weakened N-H stretching band over 2000--3000
cm$^{-1}$.  It has been revealed that such a discrepancy results from the lack
of vibrational anharmonicity and intermode couplings, since each harmonic N-H
mode leads to short N-H vibration that prevents its elongation towards HT and
the two independent N-H vibrations uncorrelate double HT pathways.  The
anharmonic and coupling impacts on the N-H vibrational bands have been
investigated from \textit{ab-initio} MD simulation with density functional
theory (DFT) in literature. The N-H vibrational bands from thermostated
classical-nuclei MD sampling multiple NVE/DFT trajectories are considerably
softened due to the success of recovering the anharmonicity and coupling with
low-energy modes, located around 2700--2900 cm$^{-1}$ for
BLYP/PW\cite{gawinkowski2012vibrations}, 2500 cm$^{-1}$ for PBE and 2750--3000
cm$^{-1}$ for B3LYP-vdW functional\cite{litman2019elucidating}.  However, the
broad N-H signature extending in the lower frequency range is still missing from
DFT/MD simulation, and was recently suggested to ascribe to protonic quantum
effects based on ring-polymer MD results.  Here, we present an alternative
interpretation from 10 ps classical-nuclei \textit{ab-initio} MD/NVE simulation
using MBE(3)-OSV-MP2 correlated model, which yields VDOS spectrum that retrieves
both broadened low- and high-energy N-H stretching peaks centered at 2600
cm$^{-1}$  and 3000 cm$^{-1}$ (Figure~\ref{fig:f_vdos_por}), respectively,
by propagating classical protons. 

The MBE(3)-OSV-MP2/MD parallel simulation was carried out on 96 CPUs using an
initial porphycene structure from MBE(3)-OSV-MP2/6-31g* optimization. On the
average, about 32 seconds (9 seconds for RHF and 23 seconds for MBE(3)-OSV-MP2
energy and gradient evaluations) were spent in each MD step, and the entire
20000 MD steps were completed in less than 8 days. The RDF for N-N distances
(Figure~\ref{fig:f_rdf_por}) shows that there are two broad peaks between 2.5
and 3.0 Angstrom in which the first peak resembles the signature of the nitrogen
pairs involved in proton transfers causing the respective $trans$-Pc and
$cis$-Pc tautomerization. As shown in Figure~\ref{fig:f_rdf_por}, the peak
position is assigned to the N-N pair of $trans$-Pc, while the lower peak
shoulder is given to the shorter N-N pair of $cis$-Pc, suggesting more
$trans$-Pc.  The VDOS spectrum also features the weak $cis$-Pc band centered at
2600 cm$^{-1}$ and relatively stronger $trans$-Pc band centered at 3000
cm$^{-1}$, which also indicates more $trans$-Pc than $cis$-Pc. In contrast, the
literature ring-polymer DFT/MD results however concluded a larger proportion of
$cis$-Pc tautomer and thus stronger hydrogen bonds, leading to more hydrogen
transfer due to the inclusion of quantal protons. More detailed studies
combining both MP2-level correlated electrons and quantal protons are therefore
desired.

\section{CONCLUSIONS}

We have developed a low-order scaling and parallel algorithm for MBE(3)-OSV-MP2
analytical gradient computation of large molecules using the third-order
many-body expansion of OSV-MP2 wavefunction and density matrices.  By
construction, each 1b cluster is orbital-specific to an LMO and computed in the
corresponding basis of OSVs, leading to the linearly increasing number of 1b
clusters with molecular sizes; the local 2b and 3b clusters are respectively
specific to orbital pair and triple correlations and screened to achieve the
effective linear growths. Higher-order MBE($n$) ($n>3$) clusters are shown
insignificant to both correlation energies and analytical gradients, and can be
entirely neglected. By further introducing  correlation approximations to
long-range 2b clusters and by exploiting the sparsity in density fitting
integrals and OSV relaxation vectors, the computational costs are mitigated to
the low-order scaling in the linear and quadratic regimes for correlation energy
and gradient, respectively. Moreover, by managing the global and local data
arrays of selected MBE(3)-OSV-MP2 intermediates in the hybrid RMA and SHM
parallelism through passive one-sided communication across multiple nodes, the
highly parallelized algorithm conveys an implementation that enables fast and
scalable MBE(3)-OSV-MP2 computations of energy and analytical gradient using a
large number of CPUs. The computations of $\mathbf{Y}_i$ intermediate and
Z-vector equation remain the main components to the overall runtime cost to
obtain accurate analytical gradients of large molecules of a few hundred atoms.
Existing techniques can be envisaged to improve the scalability of these steps.
For instances, multipole-based long-range integrals accounting for the asymptotic
behaviour of $1/R$ can be utilized to expeditiously estimate two-electron
potential in Z-vector equation.  Moreover, fast evaluation of two-center or
multi-center molecular integrals and their derivatives are recently available
~\cite{peels2020molecular,peels2020fast}.  The sparse fittings have not yet been
implemented for accelerating evaluation of the product $\mathbf{Y}_i^\dagger
\mathbf{J}_p$.

The correlation energies of Baker testing molecules based on the normal
selection of MBE(3) clusters are recovered by $>99.85\%$ and the gradient RMSDs
are $< 3\times 10^{-4}$ au from those of canonical RI-MP2, which are mostly
comparable to the original OSV-MP2 results using the same OSV set. The optimized
structures have been benchmarked for medium size molecules containing second and
third row elements, using up to two thousand basis functions and five thousand
fitting functions.  The selections of OSVs ($l_\text{osv}=10^{-4}$), 2b clusters
($l_\text{2b}=10^{-2}$) and 3b clusters ($l_\text{3b}=0.1$) yield small
MBE(3)-OSV-MP2 errors of 0.1--0.5 pm for short bonded interatomic distances and
1.1--1.5 pm for long non-bonded distances, using only a small fraction of 2b and
3b clusters. The NVE MD simulations of protonated water tetramer and hexamer
driven by MBE(3)-OSV-MP2 gradients have been performed, and the resulting RDF
and VDOS spectra are in excellent agreement to the normal OSV-MP2 benchmarks.
The efficiencies and capabilities of the MBE(3)-OSV-MP2 gradient implementation
were further demonstrated in parallel computations of C$_{60}$@catcher (148 atoms)
and (H$_2$O)$_{190}$ (570 atoms) molecules on 24 CPUs, with the total
runtime of about 2.7 and 46 hours in a single gradient step with def2-tzvp and
cc-pvtz basis sets, respectively. Finally, in two brief applications, we show
that the MBE(3)-OSV-MP2 algorithm permits the differentiation of the subtle
structure changes in interlocked Cu-catenane supramolecule by varying ligand
length, and also 10 picoseconds long MD simulation of porphycene ($\sim40$
atoms) that reveals N-H stretching signature associated with inter-convertible
tautomers.

\begin{acknowledgement}

The authors acknowledge financial supports from the Hong Kong Research Grant
Council (Grant No. ECS27307517 and GRF17309020). We are grateful to the
Computational Initiative provided by the Faculty of Science at the University of
Hong Kong and Tianhe-2 computing service at the National Supercomputer Center in
Guangzhou of China for their technical supports and allocation of CPU hours.
J.Y. acknowledges the research program of AIR@InnoHK cluster from the Innovation
and Technology Commission of Hong Kong SAR of China.  Q.L. thanks Ruiyi Zhou for
discussions.

\end{acknowledgement}

\begin{suppinfo}

The file Supporting supporting.pdf contains further results 
of the computations and is available free of charge.

\end{suppinfo}


\bibliography{manuscript}

\appendix

\section*{Appendix}
\section*{\label{app:pseudo}Gradient implementation}

\vspace{-1em}
\begin{algorithm}[H]
\caption{\label{alg:pseudo}Pseudocode and parallelism for MBE(3)-OSV-MP2 analytical gradients.}
\begin{algorithmic}
  \State{\textbf{***MBE(3)-OSV-MP2 energy***}}
  \State{Parallel tasks on AO shell pairs $(\alpha\beta)$ for $(\alpha\beta|A)$:}
    \State{\quad sparse fitting $\mathbf{J}_i$ and $\mathbf{T}_{kk}$, placed on shared disk}
  \State{Parallel tasks on occupied LMOs:}
    \State{\quad generation of exact OSVs or ID-OSVs}
  \State{Parallel tasks on $ij$ pairs in RMA:}
   \State \quad evaluate OSV overlap $\mathbf{S}_{(i,j)}$ and $\mathbf{F}_{(i,j)}$;
   \State \quad 2b and 3b clustering and sorting;
   \State \quad evaluate $\mathbf{K}_{(ij,ij)}$ and  $\mathbf{K}_{(i,j)}$.
  \State{Parallel tasks on 1b/2b/3b clusters:}
    \State \quad solve MBE(3)-OSV-MP2 residual equations;
    \State \quad collect $\mathbf{T}_{(ij,ij)}$ and  $\mathbf{R}_{(ij,ij)}$ in RMA.
  \State{\textbf{***MBE(3)-OSV-MP2 gradient***}}
  \State{Parallel tasks on 1b/2b/3b clusters:}
    \State \quad $\mathbf{D}^\text{1b},~\mathbf{D}^\text{2b}~\text{and}~\mathbf{D}^\text{3b}$, $\mathbf{N}_{ij}^\text{1b},~\mathbf{N}_{ij}^\text{2b}~\text{and}~\mathbf{N}_{ij}^\text{3b}$;
    \State \qquad accumulate $\mathbf{D}$ in SHM assigned to root process; 
    \State \qquad one-sided accumulation for $\mathbf{N}_{ij}$ in RMA.
  \State{Parallel tasks on LMO $i$ batches:}
    \State{\quad local loop on $j$ for each task $i$:}
      \State \qquad $\mathbf{Y}_i\leftarrow\mathbf{Y}_i+\mathbf{J}_{j}(\mathbf{Q}_{i}\mathbf{Q}_{j})\mathbf{\overline{T}}_{ij,ij}(\mathbf{Q}_{i}\mathbf{Q}_{j})^{\dagger}$; 
      \State \qquad $\mathbf{N}_i \leftarrow \mathbf{N}_i+\mathbf{N}_{ij}+\overline{\mathbf{T}}_{(ij,ij)}\mathbf{K'}_{(ij,ij)}$.
    \State \quad $\mathbf{X}_{i} = [\mathbf{Q}_{i}(\mathbf{N}_{i}\Delta\mathbf{G}_{ii})\mathbf{Q}'^{\dagger}_{i}]/(f_{aa}+f_{bb}-2f_{ii})$;
    \State \quad $\mathbf{Y}_i \leftarrow \mathbf{Y}_i +\mathbf{J}_{i}(\mathbf{X}_{i} + \mathbf{X}^{\dagger}_{i})$, stored on shared disk.
  \State{Parallel tasks on batches $A$ for $(\alpha\beta|A)$ and $(\alpha\beta|A)^{(\lambda)}$:}
    \State \quad $\mathbf{y}(A) + = \mathbf{y}(A)$; $\mathbf{y'}(A)+ =\mathbf{y'}(A)$;
                 collect $\sum_i\braket{\mathbf{P}_{v}\mathbf{Y}_{i}^{\dagger}\mathbf{J}_{i}^{(\lambda)}}$.
  \State Evaluate $\sum_i \mathbf{P}_{v}\mathbf{Y}_{i}^{\dagger}\mathbf{J}_{i}$ and $\braket{\mathbf{P}_{v} \mathbf{Y}_{i}^{\dagger} \mathbf{J}_p}$ on root process.
  \State \textbf{***OSV Z-vector***}
  \State{Parallel on AO shell pairs $(\alpha\beta)$ for RHF $(\alpha\beta|A)$:}
    \State \quad $(i\beta\vert A)=\sum_{\alpha\beta}(\alpha\beta\vert A)C_{\alpha i}$; 
                 $(\tilde{i}\beta\vert A)=\sum_{\alpha\beta} (\alpha\beta\vert A)Z_{\alpha i}$.
  \State{Parallel tasks on LMO $i$ batches:}
    \State \quad sparse fittings for $J_{i,A'\alpha}$, $J_{i,A'j}$ and $J''_{i,A''j}$; 
    \State \quad update $\Gamma_{A'}=\sum_{\alpha i}J_{i,A'\alpha}Z_{\alpha i}$;
    \State \quad 
      $K_{\alpha i}\leftarrow K_{\alpha i}+\sum_{jA'}J_{j,A'\alpha}J_{i,A'j}+\sum_{jA''}(\tilde{i}\alpha\vert A'')J''_{i,A''j}$; 
    \State \quad $J_{\alpha i}\leftarrow J_{\alpha i}+\sum_{A'}J_{i,A'\alpha}\Gamma_{A'}$; 
    \State \quad $U_{ai}+=4J_{\alpha i}-K_{\alpha i}$.
  \State Solve OSV Z-vector equation.
\end{algorithmic}
\end{algorithm}

\end{document}